\documentclass[vecphys]{svmult}

\usepackage{natbib}
\usepackage{makeidx,graphicx,lscape,longtable,url,threeparttable,rotating,amsmath}

\usepackage{multicol}        
\usepackage[bottom]{footmisc}

\makeindex             

\begin{document}

\newcommand{\arpc}{{\em Ann. Rev. Phys.  Chem. }}

\newcommand\lesssim{\mathrel{\hbox{\rlap{\hbox{\lower4pt\hbox{$\sim$}}}\hbox{$<$
}}}}
\newcommand\gtrsim{\mathrel{\hbox{\rlap{\hbox{\lower4pt\hbox{$\sim$}}}\hbox{$>$
}}}}

\newcommand\ion[2]{#1$\;${\scshape{#2}}}
\newcommand{\nat}{{\em Nature }}
\newcommand{\aap}{{\em Astron. \& Astrophys. }}
\newcommand{\gca}{{\em Geochim. Cosmochem. Acta}}
\newcommand{\aj}{{\em Astron.~J. }}
\newcommand{\apj}{{\em Astrophys.~J. }}
\newcommand{\araa}{{\em Ann. Rev. Astron. Astrophys. }} 
\newcommand{\apjl}{{\em Astrophys.~J.~Letters }}
\newcommand{\apjs}{{\em Astrophys.~J.~Suppl. }}
\newcommand{\jcp}{{\em J. Chem. Phys. }}
\newcommand{\jgr}{{\em J. Geophys. Res}}
\newcommand{\apss}{{\em Astrophys.~Space~Sci. }}
\newcommand{\icarus}{{\em Icarus }}
\newcommand{\mnras}{{\em MNRAS }}
\newcommand{\pasp}{{\em Pub. Astron. Soc. Pacific }}
\newcommand{\pasj}{{\em Pub. Astron. Soc. Japan }}
\newcommand{\planss}{{\em Plan. Space Sci. }}
\newcommand{\physrep}{{\em Phys. Rep.}}
\newcommand{\prb}{{\em Phys. Rev. B}}
\newcommand{\bain}{{\em Bull.~Astron.~Inst.~Netherlands }}

\newcommand{\cc}{\mbox{cm$^{-3}$}}
\newcommand{\tauv}{\mbox{$\tau_V$}}
\newcommand{\av}{\mbox{$A_V$}}
\newcommand{\ra}{\mbox{$\rightarrow$}}
\newcommand{\nhtwo}{\mbox{n$_{H_{2}}$}}

\def\HI{H{\smc I}}
\def\HII{H{\smc II}}
\def\m17{M~17}               
\def\cepa{Cepheus~A}               
\def\Htwo{H$_2$}               
\def\HtwoO{H$_2$O}             
\def\orthoHtwoO{o--H$_2$O}             
\def\HtwoeiO{H$_2^{18}$O}             
\def\HtwoCO{H$_2$CO}           
\def\HtwoCS{H$_2$CS}           
\def\Hthreep{H$_3^+$}          
\def\HtwoDp{H$_2$D$^+$}        
\def\DtwoHp{D$_2$H$^+$}        
\def\HthreeOp{H$_3$O$^+$}          
\def\Dthreep{D$_3^+$}          
\def\HCOp{HCO$^+$}             
\def\DCOp{DCO$^+$}             
\def\HthCOp{H$^{13}$CO$^+$}    
\def\HtwCsiOp{H$^{12}$C$^{16}$O$^+$} 
\def\HCSp{HCS$^+$}             
\def\HthCN{H$^{13}$CN}         
\def\HCfiN{HC$^{15}$N}         
\def\HtwCfoN{H$^{12}$C$^{14}$N}  
\def\HNthC{HN$^{13}$C}         
\def\HfoNtwC{H$^{14}$N$^{12}$C}  
\def\HCthreeN{HC$_3$N}         
\def\twCO{$^{12}$CO}           
\def\thCO{$^{13}$CO}           
\def\CseO{C$^{17}$O}           
\def\CeiO{C$^{18}$O}           
\def\twCsiO{$^{12}$C$^{16}$O}  
\def\thCsiO{$^{13}$C$^{16}$O}  
\def\twCeiO{$^{12}$C$^{18}$O}  
\def\thCeiO{$^{13}$C$^{18}$O}  
\def\CtfS{C$^{34}$S}           
\def\thCS{$^{13}$CS}           
\def\twCttS{$^{12}$C$^{32}$S}  
\def\tfSO{$^{34}$SO}           
\def\ttSsiO{$^{32}$S$^{16}$O}  
\def\SOtwo{SO$_2$}             
\def\tfSOtwo{$^{34}$SO$_2$}    
\def\SiO{SiO}             
\def\Ntwo{N$_2$}               
\def\Otwo{O$_2$}               
\def\NtwoHp{N$_2$H$^+$}        
\def\NtwoDp{N$_2$D$^+$}        
\def\NHthree{NH$_{3}$}         
\def\CHthreeCCH{CH$_3$C$_{2}$H}     
\def\CHthreeCN{CH$_3$CN}       
\def\CHthreeOH{CH$_3$OH}       
\def\CHfour{CH$_4$}       
\def\COtwo{CO$_2$}       
\def\thCHthreeOH{$^{13}$CH$_3$OH}       
\def\twCHthsiOH{$^{12}$CH$_3$$^{16}$OH} 
\def\CtwoH{C$_2$H}             
\def\CtwoS{C$_2$S}             
\def\CHp{CH$^{+}$}             
\def\Cp{C$^+$}             
\def\Hp{H$^+$}             
\def\Hep{He$^+$}             
\def\CthreeHtwo{C$_3$H$_2$}    
\def\Jthoh{$J = 3/2 \to 1/2$}
\def\Johoh{$J = 1/2 \to 1/2$}
\def\Jtwel{$J = 12 \to 11$}
\def\Jelt{$J = 11 \to 10$}
\def\Jtn{$J = 10 \to 9$}
\def\Jne{$J = 9 \to 8$}
\def\Jes{$J = 8 \to 7$}
\def\Jss{$J = 7 \to 6$}
\def\Jsf{$J = 6 \to 5$}
\def\Jff{$J = 5 \to 4$}
\def\Jft{$J = 4 \to 3$}
\def\Jtt{$J = 3 \to 2$}
\def\Jto{$J = 2 \to 1$}
\def\Joz{$J = 1 \to 0$}
\def\WCO{W({\rm CO})}
\def\Wtw{W({\rm ^{12}CO})}
\def\Wth{W({\rm ^{13}CO})}
\def\dv{\Delta v}
\def\dvtw{\Delta v({\rm ^{12}CO})}
\def\dvth{\Delta v({\rm ^{13}CO})}
\def\NCO{N({\rm CO})}
\def\Nth{N({\rm ^{13}CO})}
\def\Ntw{N({\rm ^{12}CO})}
\def\NtwCsiO{N({\rm ^{12}C^{16}O})}
\def\NthCO{N({\rm ^{13}CO})}
\def\NthCsiO{N({\rm ^{13}C^{16}O})}
\def\NtwCeiO{N({\rm ^{12}C^{18}O})}
\def\intCO{\int T_R({\rm CO})dv}
\def\inttwCsiO{\int T_R({\rm ^{12}C^{16}O})dv}
\def\intthCsiO{\int T_R({\rm ^{13}C^{16}O})dv}
\def\inttwCeiO{\int T_R({\rm ^{12}C^{18}O})dv}
\def\NHtwo{N({\rm H_2})}
\def\Wtw{W_{12}}
\def\Wth{W_{13}}
\def\kappanu{\kappa_{\nu}}
\def\phinu{\varphi_{\nu}}
\def\taunu{\tau_{\nu}}
\def\dv{\Delta v}
\def\dvFWHM{\Delta v_{FWHM}}
\def\vLSR{v_{LSR}}
\def\Rsol{R_\odot}
\def\Msol{M_\odot}
\def\MMsol{\ts 10^6\ts M_\odot}
\def\MCO{M_{\rm CO}} 
\def\Mvir{M_{\rm vir}}
\def\TAstar{T^*_A}
\def\TAstartwCO{T^*_A(^{12}{\rm CO})}
\def\TAstarthCO{T^*_A(^{13}{\rm CO})}
\def\TAstarCeiO{T^*_A({\rm C}^{18}{\rm O})}
\def\TRstar{T^*_R}
\def\TexCO{T_{ex}({\rm CO})}
\def\Trms{T_{rms}}
\def\d{^\circ}
\def\h{^{\rm h}}
\def\mi{^{\rm m}}
\def\s{^{\rm s}}
\def\mum{\ts \mu{\rm m}}
\def\mm{\ts {\rm mm}}
\def\cm{\ts {\rm cm}}
\def\percm{\ts {\rm cm}^{-1}}
\def\m{\ts {\rm m}}
\newcommand\kms{\rm{\, km \, s^{-1}}}
\def\K{\ts {\rm K}}
\def\Kkms{\ts {\rm K\ts km\ts s^{-1}}}
\def\kHz{\ts {\rm kHz}}
\def\MHz{\ts {\rm MHz}}
\def\GHz{\ts {\rm GHz}}
\def\pc{\ts {\rm pc}}
\def\kpc{\ts {\rm kpc}}
\def\Mpc{\ts {\rm Mpc}}
\def\cmsq{\ts {\rm cm^2}}
\def\pcsq{\ts {\rm pc^2}}
\def\dsq{\ts {\rm deg^2}}
\def\debye{\ts10^{-18}\ts {\rm esu}\ts {\rm cm}}
\def\swash2o{$1_{10} - 1_{01}$}             

\let\ap=\approx
\let\ts=\thinspace

\title*{The Chemical Evolution of Protoplanetary Disks}
\author{Edwin A. Bergin\inst{1}}
\institute{Department of Astronomy, University of Michigan, Ann Arbor MI 48109
\texttt{ebergin@umich.edu}}

%
%
\maketitle

\section{Introduction}

The origins of planets, and perhaps life itself, is intrinsically linked to the chemistry of planet formation.
In astronomy these systems are labelled as protoplanetary disks - disks on the incipient edge of planet formation.  For our Sun, the rotating  ball of gas and dust that collapsed to a disk has been called the Solar Nebula.   In this chapter we will attempt to explore the chemistry of planet-forming disks from the perspective of knowledge gained from decades of solar system study, combined with our rapidly growing knowledge of extra-solar protoplanetary disks.   This chapter is not written in the form of a review.
Rather we survey our basic knowledge of chemical/physical processes and the various  techniques that are applied to study solar/extra-solar nebular chemistry.    As such our reference list is limited to works that provide a direct discussion of the topic at hand.

A few aspects of general terminology and background are useful.  
I will use $n_{\rm H}$ to refer to the space density (particles per cubic centimeter) of atomic hydrogen, $n_{\rm H_2}$ for the molecular hydrogen density, and $n$ to the total density ($n_{\rm H}$ $+$ $n_{\rm H_2}$).  Similar terminology will be used for the column density (particles per square centimeter) which is denoted by $N$.   In addition $T_{gas}$ will refer to the gas temperature and $T_{dust}$ for the dust/grain temperature which are not always equivalent and have separate effects on the chemistry.  At high densities, where they are equivalent,  I will simply refer to the temperature, $T$.

Stars are born in molecular clouds with typical densities of a few thousand H$_2$ molecules per cubic centimeter, gas temperatures of $\sim 10 - 20$~K, and $10^3 - 10^5$ M$_{\odot}$.   These clouds exist over scale of tens of parsecs  (1 pc $= 3 \times 10^{18}$~cm), but exhibit definite substructure with stars being born in denser ($n > 10^5$ \cc ) cores with typical sizes of 0.1 pc.   The chemistry of these regions is dominated by reactions between ions and neutrals, but also by the freeze-out of molecules onto the surfaces of cold dust grains. 
A summary of the properties of the early stages can be found in the reviews of \citet{bt_araa} and \citet{difran_ppv}. 
Molecular cloud cores are rotating and, upon gravitational collapse, the infalling envelope flattens to a disk.    This stage is labeled as Class 0.  As the forming star begins to eject material the envelope is disrupted along the poles and the star/disk system begins the process of destroying its natal envelope.  In this ``Class I'' stage the star accretes material from the disk and the disk accretes from the envelope
\citep[see ][for a discussion of astronomical classification]{als87, awb93}.
As the envelope dissipates, the system enters the Class II stage wherein material accretes onto the star from the disk.  The disk surface is directly exposed to energetic UV radiation and X-rays from the star, but also the interstellar radiation field.    Observational systems in the Class II stage are often called T Tauri stars.
These disks have strong radial and vertical gradients in physical properties with most of the mass residing in the middle of the disk, labeled as the midplane. 
  It is the midplane that is the site of planet formation.   
 The density of the midplane significantly exceeds that of the dense natal core, by many orders of magnitude.
  The midplane is in general, colder than the disk upper layers, which can also be called the disk atmosphere. Finally, the outer regions of the disk (r $>$ 10 AU) has reduced pressure and temperature, but contains most of the mass.   \citet{bergin_ppv} and \citet{cc_messii} provide recent summaries of disk chemical evolution.   This chapter will provide greater detail on the methodology and techniques used to explore the chemistry of protoplanetary disks and the reader is referred to the previous reviews for additional information regarding both observations and theory.

We will start our study by discussing some key  observational results, beginning with bodies in our solar system  and extending to circumstellar disk systems at a typical distance of 60--140~pc.  We expand our exploration to the basic facets of theoretical studies of disk chemistry: thermodynamic equilibrium and chemical kinetics.   This is followed by an outline of our knowledge of key physical processes. 
In the final sections, we synthesize these results to summarize our theoretical understanding of 
 the chemistry of protoplanetary disks

\section{Observational Constraints}
\label{sec:obs}

There exists a wealth of data on the chemical evolution of our own solar system regarding the composition of planetary bodies, moons, asteroids, comets, and meteorites.  In addition, there is a rapidly growing observational sample of molecular and atomic transitions in extrasolar protoplanetary disks.  The current astronomical dataset is limited only to a few objects which have been subject to deep searches.  However, the upcoming Atacama Large Millimeter Array (ALMA) will dramatically expand our current capabilities and offers great promise for gains in our understanding of planet formation.  This will directly complement knowledge gained from the remnants and products of planet formation within the  solar system.
For more detailed views of disk chemistry the reader is referred to the recent compilations found in Meteorites and the Early Solar System II \citep{messii}, the Treatise on Geochemistry \citep[Vol. 1,][]{mcp05}, and Protostars and Planets V \citep{ppv}.

\subsection{Planets, Comets, and Meteorites: Tracing the Midplane}

This paper cannot do justice to the decades of study gained in the study of bodies within our solar system.  
Instead we will highlight a few basic facets of our understanding and  a few examples where the chemical composition is a fossil remnant that tracks conditions which existed billions of years ago.

\subsubsection{Planets}

\begin{figure}
\centering
\includegraphics[height=17.cm, angle=0]{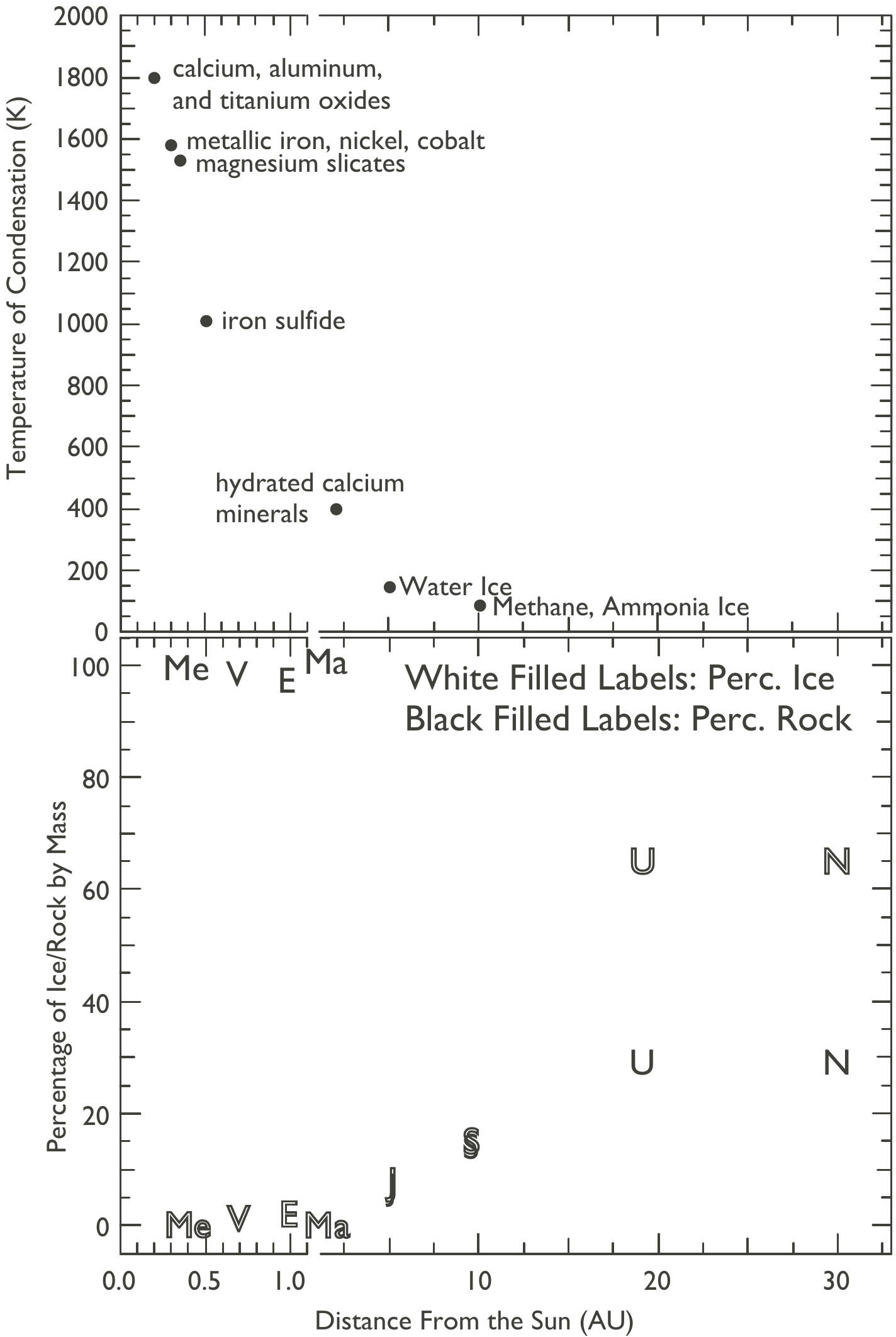}
\caption{\small
{\em Top:} Plot of temperature of condensation for materials that represent some of the major contributors to planetary composition as a function of distance from the Sun.  This part of the figure is adapted from NASA's Genesis mission educational materials.
Condensation temperature and estimated position from \citet{lewis90}.
{\em Bottom:}
Plot of percentage of planetary mass constituted by rocks (black filled labels) and
ices (white filled labels).  Labels represent each of the 8 major planets in the solar system.  Jupiter, Saturn, Uranus, and Neptune estimates are from \citet{guillot99a} and \citet{guillot99b}.  Values for the giant planets have larger uncertainties than terrestrial planets.
}
\label{fig:ss}
\end{figure}

The standard model for the formation of the Earth and other rocky planets is that
the temperature of the nebula was initially hot enough that all primordial grains sublimated.  As the nebula cooled various species condensed out of the hot gas depending on their condensation temperature (defined in \S~\ref{sec:conden}).  The nebula also had strong radial temperature gradients such that different species could first condense at larger radii.     This model and various issues are summarized in \citet{palme_mess} and \citet{davis_messii}.   
Fig.~\ref{fig:ss} shows the condensation temperatures of various minerals and ices along with the radii where these species potentially condensed in the nebula as estimated by \citet{lewis90}.   The fraction of rock/ices in planets is also provided on this plot and clearly follows this trend.   It is a basic astronomical fact that the rocky planets reside in the inner nebula, with gas giants in the outer; however, this is supplemented by  Uranus and Neptune having incorporated a greater percentage of ices.  
   Thus, basic molecular properties (i.e. temperature of condensation) with the nebular thermal structure work in tandem to determine the composition of bodies in the nebula to be supplemented by  dynamical evolution \citep[e.g.,][]{morby00}.   Other facets of chemical composition of bodies in the solar system in relation to nebular thermal structure are discussed by \citet{lewis74}, providing some of the basis for the condensation temperature/radii placement in Fig.~\ref{fig:ss}.

\subsubsection{Meteorites}
\label{sec:met}

\begin{figure}[!t]
\centering
\includegraphics[height=8.5cm, angle=0]{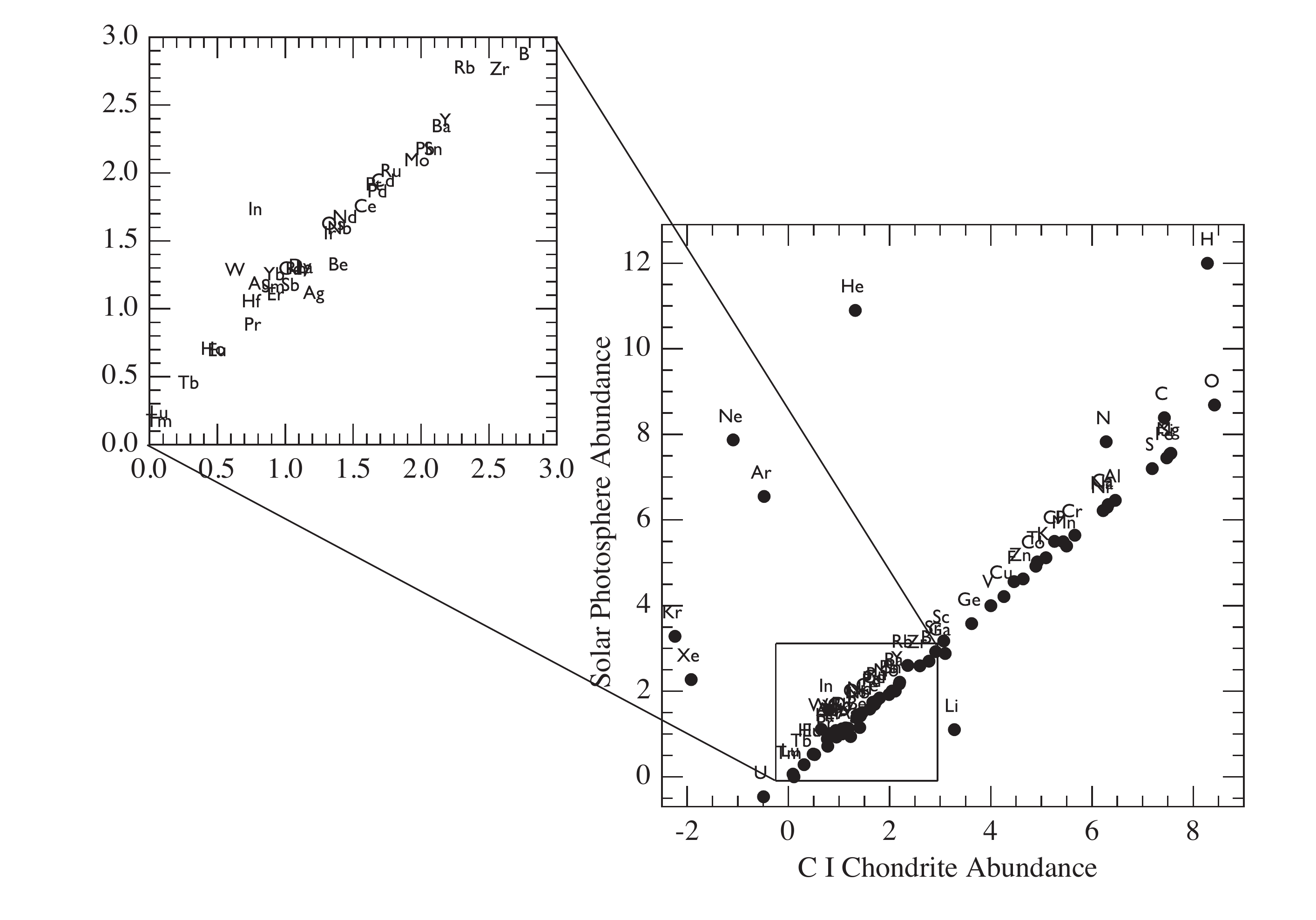}
\caption{\small
Plot of abundances measured in C I Chondrites against values estimated in the Solar Photosphere using compilation of \citet{2003ApJ...591.1220L}.
}
\label{fig:ci}
\end{figure}

Meteorites are the most directly accessible primitive material in the solar system.  Meteorites come in a number of classes, based, in part, on whether the material is undifferentiated and therefore more primitive, or part of the mantle or core of some larger planetesimal \citep[see][]{krotmcp}.   The most primitive meteorites are the carbonaceous chondrites (labelled as C).   Amongst the various types, the CI carbonaceous chondrite contains a composition that is quite similar to the Sun \citep{pj03}.  In fact in many instances abundances from CI chondrites provide a better value for the Solar composition than spectroscopy of the solar photosphere \citep{pj03, 2003ApJ...591.1220L}.  This is illustrated in Fig.~\ref{fig:ci} where we plot abundances measured in CI chondrites against those estimated in the Sun.   
 
 The exceptions to the general  agreement in Fig.~\ref{fig:ci} are telling.
Li is under-abundant in the Sun due to convective mixing into layers where Li is destroyed. 
Li is inherited from the interstellar medium and stars begin their evolution with abundant Lithium.  Primordial Li is gradually destroyed \citep{boden65, skum72}; thus the presence of Li in a stellar photosphere is generally taken as a sign of youth.
The most volatile elements have low abundances in chondrites, this includes the noble gases, along with H, N, C, and to some extent O.   Unreactive noble gases reside for the most part in the nebular gas along with H (which was in molecular form).   Nobel gases can be trapped inside the lattices of rocky and icy planetesimals and their abundances provide key clues to the evolution of terrestrial planet atmospheres \citep{pepin91, ob95, 2003Icar..165..326D}.
In the case of C and N, thermodynamic equilibrium would place these molecules in the form of CO/CH$_4$ or N$_2$/NH$_3$ (depending on the  pressure/temperature).  Thus
the temperature  in the nebula where these rocks formed must have been above the sublimation temperature for these molecules.
In the case of \ion{O}{i}, the oxygen not consumed by refractory material and CO likely resides in the form of gas-phase H$_2$O.

\begin{figure}[!t]
\centering
\includegraphics[height=10.cm, angle=0]{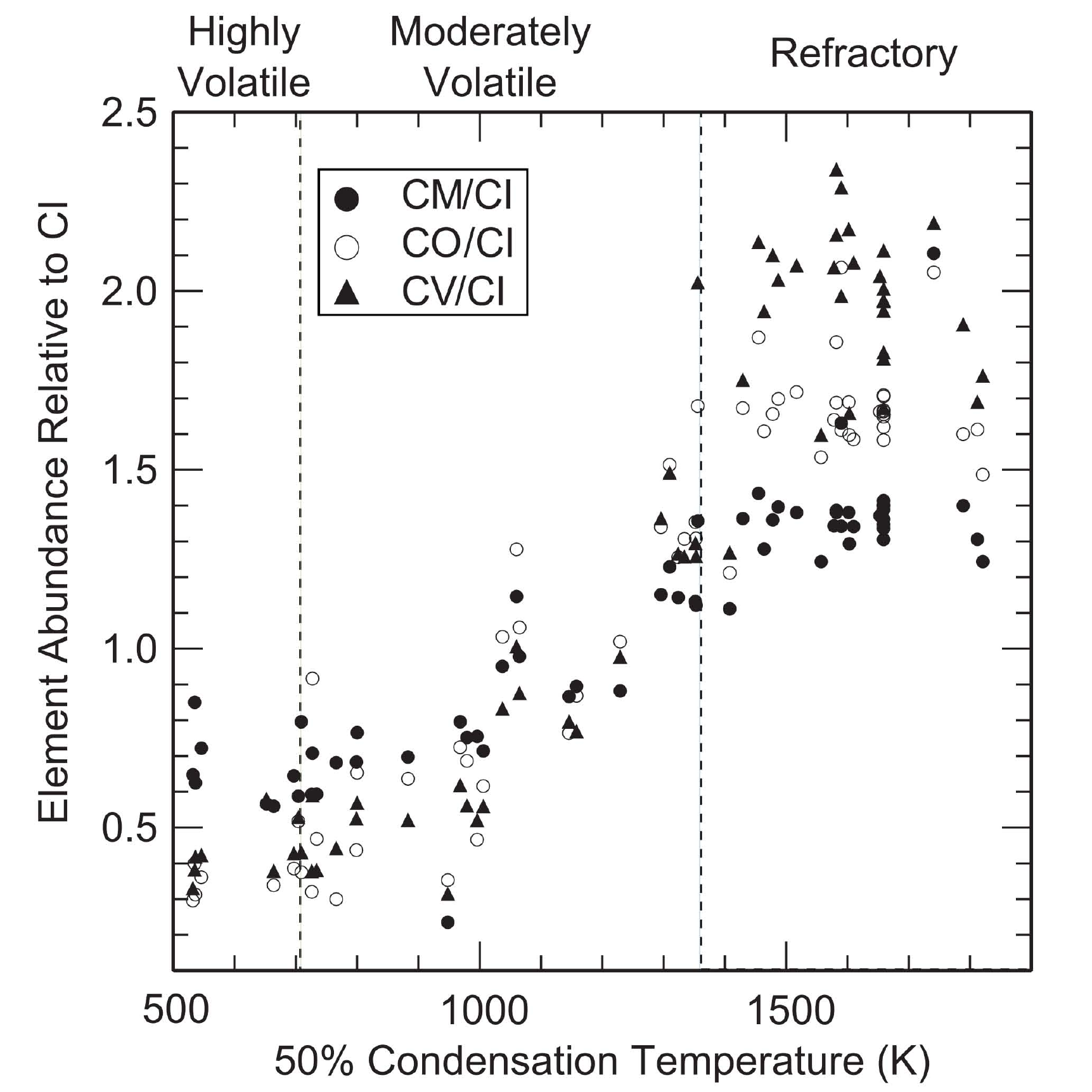}
\caption{\small
Plot of elemental abundances measured in the CM/CO/CV chondrites
normalized to CI shown as a function of the 50\% condensation temperature for that
element.   Figure adapted from \citet{abc01}.   Elemental abundances within
chondrites are taken from \citet{wk88} with condensation temperatures from
\citet{2003ApJ...591.1220L}.
}
\label{fig:tcon}
\end{figure}

An additional clue to the chemistry of rock formation in the inner nebula lies in the relative abundances of elements within the various classes of meteorites.  Fig.~\ref{fig:tcon} shows a sample of these results plotting the abundances of CM, CO, and CV carbonaceous chondrites normalized to CI as a function of condensation temperature.   For a more expanded plot, the reader is referred to \citet{davis_messii}.  This plot demonstrates that elements with higher condensation temperatures are 
more likely to be incorporated into planetesimals.  Hence,  rock formation in the solar nebula was in some sense a volatility controlled process.   This can involve aspects of both condensation in a cooling nebula and potentially evaporation during heating events \citep[][]{palme_mess, davis_messii}.

\subsubsection{Comets}

Because they harbor volatile ices,
comets are likely the most pristine remnants of the formation of the solar system, tracing conditions that existed in the outer tens of AU in the solar nebula.
Long period comets, with periods $>$ 200 yrs, originate in the Oort cloud, while short period comets
($<$ 200 yrs) are believed to originate in the Kuiper belt.   Oort cloud comets are believed to have formed predominantly in the region between Uranus and Neptune and scattered to distant orbits via interactions with the giant planets \citep{f97, dwld04}.    The chemical composition of volatiles in cometary coma can be studied via remote sensing techniques with the basic physics summarized in \S~\ref{sec:molast}.
In general,
there are some gross similarities in the abundances of volatiles in cometary coma to that of interstellar ices, which are suggestive of a direct link \citep[e.g.][]{bm00, ecw04}.   In particular water, CO, and CO$_2$, are the most abundant ices in comets and in the ISM, with smaller contributions from CH$_4$, NH$_3$, and others.  
However,  there is also a wide degree of chemical inhomogeneity in the cometary inventory.   In particular a study of 85 short-period comets by \citet{ahearn95} found that a fraction appeared to be depleted of the carbon chain  precursors that produce C$_2$.    Such diversity is not limited to short-period comets as Oort cloud comets also exhibit variations in CO, C$_2$H$_6$, and CH$_4$ relative to H$_2$O, along with other compositional differences \citep[see][and references therein]{bcmw04, disanti08}.
Methanol also stands out as another species where variations are found within the Oort cloud population of comets and also short period comets \citep{disanti08}.
These differences likely point to systematic changes in the chemistry throughout the cometary formation zone perhaps due to a diversity of formation radii for cometary nuclei.

One potential clue to the formation of cometary volatiles lies in the ortho/para ratio of water ice.
The spin pairing of the hydrogen atoms  for H$_2$O leads to two independent forms, para-H$_2$O (spins anti-parallel with total nuclear spin $I_p = 0$) and ortho-H$_2$O
(spins parallel with total nuclear spin $I_p = 1$). The lowest energy level of para-H$_2$O is $\sim 34$~K below that of ortho-H$_2$O and if water forms at temperatures below this difference then water will preferentially be in the form of para-H$_2$O.
The ortho/para ratio 
depends on the distribution of population within the energy levels that are characterized by three quantum numbers  ($J, K_+, K_-$)  used for asymmetric tops (\S~\ref{sec:molast}).
In thermodynamic equilibrium the ratio is given by \citep{mwl87}:

\begin{equation}
{\rm o/p} = \frac{2I_o + 1}{2I_p + 1} \frac{\Sigma (2J + 1)e^{-E_o(J,K_+, K_-)/kT}}{\Sigma (2J + 1) e^{-E_p(J,K_+, K_-)/kT}},
\end{equation}

\noindent
and at high temperature is set by the ratio of the spin statistical weights ($2I + 1$),  o/p $=$ 3:1.   When water forms in the gas via exothermic reactions the energy is much greater than the ground state energy difference and the ortho/para ratio reflects the 3:1 ratio of statistical weights between these species.   
When water forms on the surfaces of cold dust grains it is believed that the excess energy in the reaction is shared with the grain and the water molecules equilibrate to an ortho/para ratio at the grain temperature \citep[e.g.][]{limbach06}.     If the grain temperature is below $\sim 50$ K then the ortho/para ratio will lie below 3:1.     In Comet P/Halley, \citet{mwl87} measured an o/p ratio of 2.73 $\pm$ 0.17.   This is consistent with an equilibrium spin temperature of $\sim 30$ K, which is similar to that measured for water and other species (ammonia, methane) in other cometary coma \citep{kawakita06}.   
\citet{bonev07} provide a summary of water ortho/para ratios measured in cometary coma which span a range of spin temperatures from $\sim 20 - 40$ K.     One potential interpretation of this ratio is that cometary ice formed at temperatures near 30 K, possibly in the interstellar medium, although other explanations exist  \citep[see][for a discussion]{bcmw04}.

\subsubsection{Isotopic Chemistry}

\begin{figure}[t]
\centering
\includegraphics[height=12.cm, angle=90]{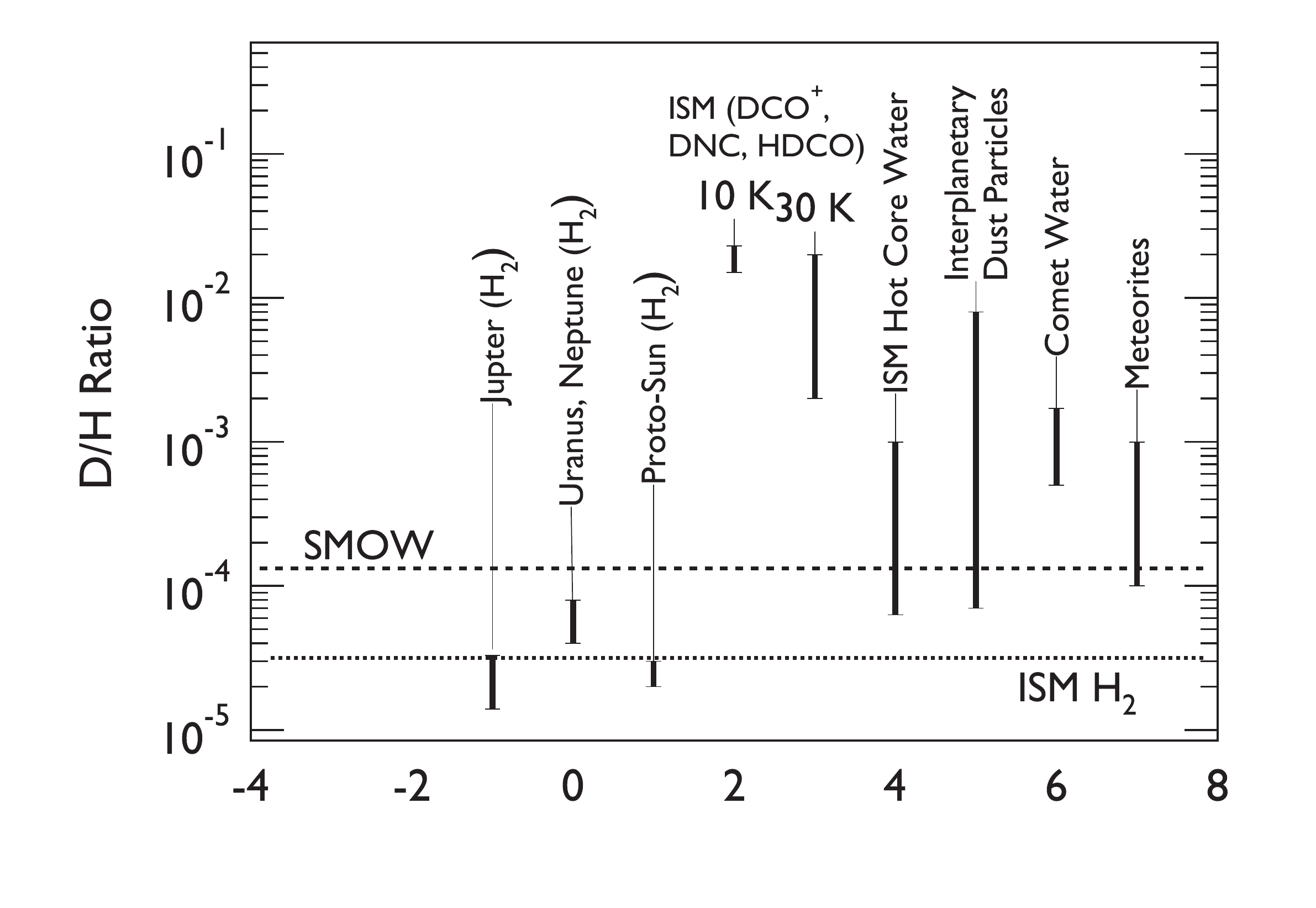}
\caption{\small
D/H ratios measured in key species in Solar System bodies and molecules in the ISM.   Plot adapted from \citet{messenger_iso} using additional ratios taken from the tabulation of  \citet{robert_messii} with references therein.
The Jupiter value is an in situ measurement from the Galileo probe \citep{mahaffy98} and ISO observations \citep{lellouch96}, while the protosolar ratio is estimated from measurements of $^3$He in the Solar wind \citep{gg98}.  SMOW -- Standard Mean Ocean Water -- is the reference standard for the Earth's ocean.    The ISM value is based upon measurements of atomic D/H in clouds within 1 kpc of the Sun \citep{wood04}.
}
\label{fig:dh}
\end{figure}

A clear chemical constraint on planet formation
is the deuterium enrichment observed in the Solar system.   In Fig.~\ref{fig:dh} we provide a compilation of (D/H) ratios seen in a variety of Solar System bodies and molecules in the ISM.   This figure essentially summarizes salient issues as determined by cosmochemical and geological studies over the past decades. (1) Gas in the early solar nebula (proto-Sun, Jupiter) is believed to have a ratio comparable to that seen in the main mass reservoir in the natal  cloud (i.e., ISM H$_2$).  (2) Moving towards bodies that formed at greater distances, there is a clear trend towards deuterium enrichment in trace gas components (Uranus, Neptune).   (3) The material that formed directly from ices and rocks (comets, IDP, meteorites) exhibit large enrichments. (4) The Earth's oceans are enriched in deuterium.
 
Based on astronomical studies of the dense ($n > 10^5$ cm$^{-3}$) cores of molecular clouds -- the sites of star and planet formation -- we know that these enrichments are readily initiated by a 
sequence of reactions that are favored for $T_{gas} \lesssim 30$ K \citep[\S~\ref{sec:dfrac};][]{gr81, millar_dfrac}.  Indeed large (D/H) ratios are observed inside molecular cloud cores, both in the gas-phase and in evaporated ices \citep[Fig.~\ref{fig:dh},][]{ceccarelli_ppv}.   The same cold ($T_{gas} \lesssim 30$ K) non-equilibrium gas chemistry that gives rise to the deuterium fractionation in molecular cloud cores is  also observed to be active in the outer reaches of extrasolar protoplanetary disks \citep[][and references therein]{bergin_ppv}.  These inferences hint that the Earth received a contribution to its water content from some cold reservoir (e.g. comets, icy asteroids).

Beyond hydrogen, meteorites also exhibit a number of other isotopic anomalies \citep[see ][for a complete discussion]{messii, mcp05}.  Of particular interest are the isotopic enhancement seen in carbon, oxygen, and nitrogen which are potentially related to kinetic chemical effects early in nebular evolution or  perhaps in the cloud that collapsed to form the Sun \citep{clayton_areps, floss04, yurimoto_ppv, rc08}.

\subsection{Astrophysical Techniques}

Astronomy offers several methods to explore the predominantly molecular chemistry of protoplanetary disks.  In general, molecules are fantastic probes of their environment.  In the following we will outline a few basic facets of molecular astrophysics and discuss the current status of observations.
A listing of molecules and transitions that have been detected in protoplanetary disks is provided in Table~\ref{tab:obs}.

\subsubsection{Molecular Astrophysics}
\label{sec:molast}

For atomic species the electronic states are discrete, or quantized, and
their emission spectra has been successfully characterized using the 
principles of quantum mechanics \citep[e.g.][or an astronomical perspective]{cowley00}.
In comparison to atoms,
molecules have more complicated structures where rotational and vibrational
motions are coupled to electronic states.  Similar to atoms, the
various motions/states are quantized.    A more complete description of the molecular physics and spectroscopy is provided by \citet{ts} and \citet[][and subsequent volumes]{herzberg_v1}.     \citet{evans_araa} provides a more descriptive outline regarding the use of molecules as astrophysical probes.

The Born-Oppenheimer approximation \citep{born_opp}
allows the separation of the nuclear and electronic
motions in the wave function.  Thus molecular transitions are characterized
by three different energies: rotational, vibrational, and electronic.
Electronic transitions have energies
typically $\sim$4 eV (40,000 K) with lines found at visible and ultraviolet (UV) wavelengths.
Vibrational levels probe energies of $\sim$0.1 eV (1,000 K) and are, for the most part, found in the near to mid infrared ($\sim 2 - 20 \mu$m).
Rotational transitions
of the ground vibrational and electronic states have
energies below 0.01 eV (100 K) and emit at millimeter and submillimeter wavelengths.
The wide range of temperatures and densities present in protoplanetary disks has led to 
detections of electronic, vibrational, and low/high energy rotational lines of a variety molecular species
 (Table~\ref{tab:obs}).

For the bulk of the disk mass, which has temperatures below tens of Kelvin, the disk is best traced by molecular rotational lines.   
Solutions of the wave equation 
for a ``rigid'' rotating linear molecule yield rotational energy levels:

\begin{equation}
E_J = \frac{\hbar ^2}{2I}J(J+1)
\end{equation}

\noindent where $I = \mu a_0^2$ is the moment of inertia ($\mu$ is the reduced mass, and $a_0$ the bond length, typically $\sim 1$ \AA )  and
$J$ is the angular momentum quantum number.   The quantity
$\hbar ^2/2I$ is redefined as the rotation constant, $B_0$, which is
an intrinsic property of a given molecule.  
For linear molecules the excitation is simplified because  
there is symmetry about all rotational axes and the selection rules for electric dipole 
transitions are $\Delta J \pm 1$.   For these molecules
the frequency of a rotational transition from level $J$ to level $J-1$ is,

\begin{equation}
\nu = E_J(J)-E(J-1) = 2B_0J.
\label{eq:nu}
\end{equation}

\noindent Heavy linear molecules with a large moment of inertia (e.g. HCN, CS, CO)
have small rotational constants and, in consequence, have 
more transitions at longer millimeter/centimeter wavelengths.  Lighter molecules, such as H$_2$ or HD, have larger energy spacings and emit primarily in the near to far-infrared ($2-115 \mu$m).   Molecules have 3 degrees of rotational freedom and the simplest case is the linear rigid rotor.
Some non-linear molecules  have  a degree of rotational symmetry.   For a symmetric top molecule two of the principal moments of inertia are equal.   In this case the rotational energy spacing is charactered by 2 quantum numbers ($J, K$).   Due to their energy level spacing and excitation properties, symmetric tops have been used as probes of the gas temperature.   Prominent examples are NH$_3$ or CH$_3$C$_2$H \citep{ht83, bgsu94}.
Non-linear molecules with no symmetry are called asymmetric tops.  The energy levels for these species are described by three quantum numbers ($J, K_+, K_-$). 
Water is the most notable species in this category, which also includes most complex organic molecules.

The spontaneous emission arises from a rotating dipole, which from an upper 
to a lower J state of a linear molecule is given by the 
Einstein A-coefficient:

\begin{equation}
A_{ul} = \frac{S}{2J_u+1}\frac{64\pi^4}{3hc^{3}}\nu^3\mu_d^2.
\label{eq:aul}
\end{equation}
 
\noindent Here $\mu_d$ is the permanent electric dipole moment of the molecule.  A common unit for the dipole moment is the Debye (1 D $= 10^{-18}$ esu cm).  $S$ is the line strength of the transition. For linear rotors $S = J_u$.

From Eqns.~\ref{eq:nu} and \ref{eq:aul},    the methodology of using molecular emission
to probe the physical environment can be demonstrated.
In a disk with density $n_{\rm H_2}$, molecules are excited through collisions with 
molecular hydrogen at a typical rate $n_{\rm H_2}$ $\gamma_{ul}$, where the collision rate
coefficient, $\gamma_{ul} \sim 10^{-11}$ cm$^{3}$ s$^{-1}$  
\citep[see][for a thorough discussion of molecular collisions]{flower03}.  
This leads to a critical density required for significant
excitation of a given transition, which is
$n_{cr} \sim A_{ul}/\gamma_{ul}$.  Using Eqn.~\ref{eq:nu} and \ref{eq:aul} the critical density 
scales as $n_{cr} \propto B_0^3\mu_d^2J^3$.   Of these, the rotation constant
and the dipole moment are molecular
properties, while $J$ depends on the observed transition.
Heavy molecules with large dipole moments, due to charge disparities, have higher critical densities and are excited in denser gas.   Thus HCN ($\mu_d = 2.98$~D) and H$_2$O ($\mu_d = 1.85$~D) are tracers of dense ($n_{\rm H_2}$ $> 10^5$ \cc ) gas, and CO ($\mu_d = 0.11$~D) traces lower density material.  Higher J transitions probe successively denser (and warmer layers), presumably at the disk surface, or  radially closer to the star.     

Molecules with identical nuclei  have no permanent electric dipole, but in some instances can emit via weaker quadrupole (H$_2$) or magnetic  dipole (O$_2$) transitions.    In the case of H$_2$, which is the dominant gas constituent, the first rotational level ($J = 2 - 0$ for a quadrupole) is 510 K above the ground  state.   The population in this state is $\propto$ exp($-E_u/kT_{gas}$), which hinders H$_2$ emission from the cold ($T_{gas} \sim 10 - 30$~K)  midplane.  Hence, similar to the ISM, other molecules with dipole moments are used as proxies to trace the molecular gas.  These species are generally heavier molecules comprised of H, C, O, and N (typically CO).   

Vibrational modes are also commonly detected in warm and dense disk systems.   Molecules with N atoms have 3N degrees of freedom.  In a linear molecule 2 degrees are rotational and 3 translational.  Thus there are 3N $-$ 5 degrees of freedom (3N $-$ 6 for non-linear).  In the case of H$_2$O there are 3 modes, two stretching and one bending.  Vibrational modes are quantized.  However, the stretching frequency can be roughly approximated in a model where the two atoms are attached via a spring.  Thus using  Hooke's law, where the vibration of spring depends on the reduced mass ($\mu_r$) and force constant $k$ (the bond),

\begin{equation}
\nu_{0} = \frac{1}{2\pi c}\sqrt{\frac{k}{\mu_r}}.
\end{equation}

\noindent  The force constant for bonds is $\sim 5 \times 10^{5}$ Dynes cm$^{-1}$ for single bonds, 10 $\times 10^{5}$ Dynes cm$^{-1}$ for double bonds,
15 $\times 10^{5}$ Dynes cm$^{-1}$ for triple bonds.  The energy of vibration depends on the vibrational quantum number, $n$:

\begin{equation}
E_{vib} = \left(n + \frac{1}{2}\right)h\nu_0.
\end{equation}

Thus heavier molecules have vibrational modes at shorter wavelengths when compared to lighter molecules.   Each vibrational state has a spectrum of rotational states called ro-vibrational transitions.
The frequency of these transitions from an upper state $'$ to a lower state $''$ is given by (ignoring some higher order terms):

\begin{equation}
\nu = \nu_0 + B_{v'} J'(J' + 1) - B_{v''}''J(J'' + 1).
\end{equation}

\noindent $B_v$ contains the rotational constant along with a correction due to the interaction between vibration and rotation (i.e. the intermolecular distance can change).  
The selection rule is $\Delta J = \pm 1$ so there are 2 cases $J' = J''+ 1$, the R branch,
and $J' = J'' - 1$, the P branch.  These are labelled as R($J$) or P($J$) where $J$ is the lower state.   In some instances $\Delta J = 0$ is allowed, which is called the Q branch.  Quadrupole transitions $\Delta J \pm 2$ are labeled as S($J$) for  $\Delta J = 2$ and O($J$) for $\Delta J = -2$.   

\subsubsection{Brief Summary of Observational Results}

Some features of the observational picture become clear in  Table~\ref{tab:obs}.   Molecular transitions that probe higher temperature gas have been found to predominantly probe the innermost regions  of the disk.  This is inferred via spatially unresolved but spectrally resolved line profiles, assuming that the line broadening is due to Keplerian rotation.  Spatially resolved observations of the lower energy rotational transitions trace the outer disk and confirm that protoplanetary disks are in Keplerian rotation \citep{ks95, sdg00} and that the temperature structure has sharp radial gradients (as expected).  In addition, analysis  of high-J molecular transitions by  \citet{vanz_etal01} and \citet{aikawa_vanz02} confirms that the surface of the irradiated disk is warmer than the midplane, as suggested by analysis of the dust spectral energy distribution \citep{calvet91, cg97}.
  It is important to state that current sub/milli-meter wave observational facilities are limited by resolution and sensitivity and, therefore, only probe size scales of tens of AU at the nearest star-forming regions (e.g. TW Hya at 65 pc, and Taurus at 140 pc).   This limitation will be significantly reduced with the advent of the ALMA array (and JWST) in the coming decade and we can expect the listing in Table~\ref{tab:obs} to drastically increase. 

\begin{threeparttable}[t]
\scriptsize
\caption{A Sample of Current Astrophysical Probes}
\begin{tabular}{l|r|l|l|l|l}
\hline\hline
\multicolumn{1}{c}{Species} &
\multicolumn{1}{c}{$\lambda (\mu$m)} &
\multicolumn{1}{c}{Transition} &
\multicolumn{1}{c}{$E_u$ (K)} &
\multicolumn{1}{c}{Radius Probed} &
\multicolumn{1}{c}{Notes\tnote{a}} \\\hline
H$_2$ & 0.10 - 0.15 & Lyman-Werner bands & 10$^5$ &r $< 1$ AU & (1) \\
H$_2$ & 2.12  & $v = 1 - 0$ S(0) & 6471	& r $\sim  10 - 40$ AU & (2)\\
CO      & 2.23 & $v = 2 - 0$ & 6300 & r $\sim 0.05 - 0.3$ AU & (4) \\
H$_2$O & $\sim 2.9$ & $v_3 = 1 - 0$ & 5000 - 10000 & r $\sim 1$ AU & (7)\\
OH   &$\sim 3$ & $v = 1 - 0$ P branch & $> 5000$ &  r $\sim 1$ AU & (7)\\
CO      & 4.6   & $v = 1 - 0$ & 3000 & r $\sim < 0.1 - 2$ AU & (5) \\
H$_2$ & 8.0 $-$ 17.0 & $v = 0 - 0$ S(1), S(2), S(4) & 1015 $-$ 3474 & r $\sim  10 - 40$ AU& (3)\\
H$_2$O & 10 $-$ 30 & $J > 4$ & $> 500$ & r $\sim 1 - 2$ AU & (6) \\
C$_2$H$_2$ & $\sim 13.7$ & $v_5 = 1 - 0$ Q branch & 1000 & $ r \sim 1$ AU & (8) \\
HCN & $\sim 14$ & $v_2 = 1 - 0$ Q branch &  1000 & $ r \sim 1$ AU & (8) \\
CO$_2$  & 14.98 & $v_2 = 1 - 0$ Q branch & 1000   & $ r \sim 1$ AU & (8)\\
\ion{Ne}{ii} & 12.81 & $^2P_{3/2} - ^2P_{1/2}$ & 1100 & r  $\sim 0.1$ AU\tnote{b} & (9) \\
CO & 460 $-$ 2600 & $6-5,3-2,2-1,1-0$ & $5 - 116$ & $r > 20$ AU\tnote{c} & (10) \\
HCO$^+$ & 1000 $-$ 3300 & $3-2, 1-0$ & $5 - 25$ & $r > 20$ AU\tnote{c} & (11) \\
CS & 1000 $-$ 3000 & $2-1, 3-2, 5-4$ & $5 - 30$ & $r > 20$ AU\tnote{c} & (12) \\
N$_2$H$^+$ & 3220 & $1-0$ & 5 &$r > 20$ AU\tnote{c} & (13)\\
H$_2$CO & 1400 $-$ 2000 & $3_{13}-2_{12}, 2_{12}-1_{11},3_{12}-2_{11}$ & 20 $-$ 32 &  $r > 20$ AU\tnote{c} & (14) \\
CN & 1000 $-$ 2500 & $3-2, 2-1$ & $5 - 30$ & $r > 20$ AU\tnote{c} & (15)\\
HCN & 850 $-$ 3300 & $4-3, 2-1, 1-0$ & $5 - 40$ & $r > 20$ AU\tnote{c} &  (16)\\
HNC & 3300 & $1-0$ & $5$ & $r > 20$ AU\tnote{c} &  (17)\\
H$_2$D$^+$ & 805 & $1_{10} - 1_{11}$ & 104 &  $r > 20$ AU\tnote{c} &  (18)\\
DCO$^+$ & 830 $-$ 1400 & $5-4, 3-2$ & 20 - 50 & $r > 20$ AU\tnote{c} &  (19)\\
DCN & 1381 & $3-2$ & 20 & $r > 20$ AU\tnote{c} &  (20)\\
\hline
\end{tabular}
\begin{tablenotes}
\item [a] References: (1) \citet{herczeg_twhya1}, (2) \citet{bwk03, bary08}, (3) \citep{bitner07},
(4) \citet{carr93}, (5) \citet{ncm03, brittain07}, (6) \citet{cn08, salyk08}, (7) \citet{salyk08},
(8) \citet{lahuis06,cn08}, (9) \citet{esp07, lahuis07, herczeg07}, (10) \citet{dutrey96, kastner_twhya, qi_phdt, vanz_etal01, qi06}, (11) \citet{dgg97, kastner_twhya, vanz_etal01, qi_chem}, (12) \citet{dgg97},
(13) \citet{dgg97, dutrey07}, (14) \citet{dgg97, tvv04}, (15) \citet{dgg97, kastner_twhya, vanz_etal01},
(16) \citet{dgg97, kastner_twhya, vanz_etal01, qi_chem}, (17)  \citet{dgg97, kastner_twhya}, (18) 
\citet{cecc_h2dp}, (19) \citet{vd_dcop, qi_chem}, (20) \citet{qi_chem}.
\item [b] If the [\ion{Ne}{ii}] emission arises from a photoevaporative wind then the emission can
arise from greater distances \citep{herczeg07}.
\item [c] It is important to note that many of these species will have rotational emission inside 20 AU, particularly in the high-J transitions.   However, the observations are currently limited by the spatial resolution, which will be overcome to a large extent by ALMA.
\end{tablenotes}
\label{tab:obs}
\end{threeparttable}
\bigskip

The chemistry of protoplanetary disks is slowly being cataloged.  Molecular emission depends on physical parameters (density, temperature, velocity field) and the molecular abundance.   
The radial distribution of density and temperature is estimated via thermal dust emission
\citep[e.g.][]{bs90}.   Power-law fits provide $\Sigma(r) \propto r^{-p}$ and $T(r) \propto r^{-q}$ with $p = 0 - 1$ and $q = 0.5 - 0.75$.   Temperatures are estimated to be $\sim 100 - 200$ K at the midplane at 1 AU with surface densities of $\sim 500 - 1500$ g cm$^{-2}$.   Modern models now take into account the disk vertical structure \citep[\S~\ref{sec:pc}, Calvet, this volume;][]{calvet91, cg97, dalessio98}.
Thus the determination of chemical composition requires  untangling the physical and chemical structure, both of which can vary in the radial and vertical direction.
 Abundances were first calculated using local thermodynamic equilibrium in a disk with radial variations in density and temperature, but now adopt more sophisticated models 
capable of exploring 2 dimensional parameter space  \citep[Monte-Carlo, Accelerated Lambda Iteration:][]{bernes79, rh91}.
 
 \begin{figure}[t]
\includegraphics[width=9cm, angle=0]{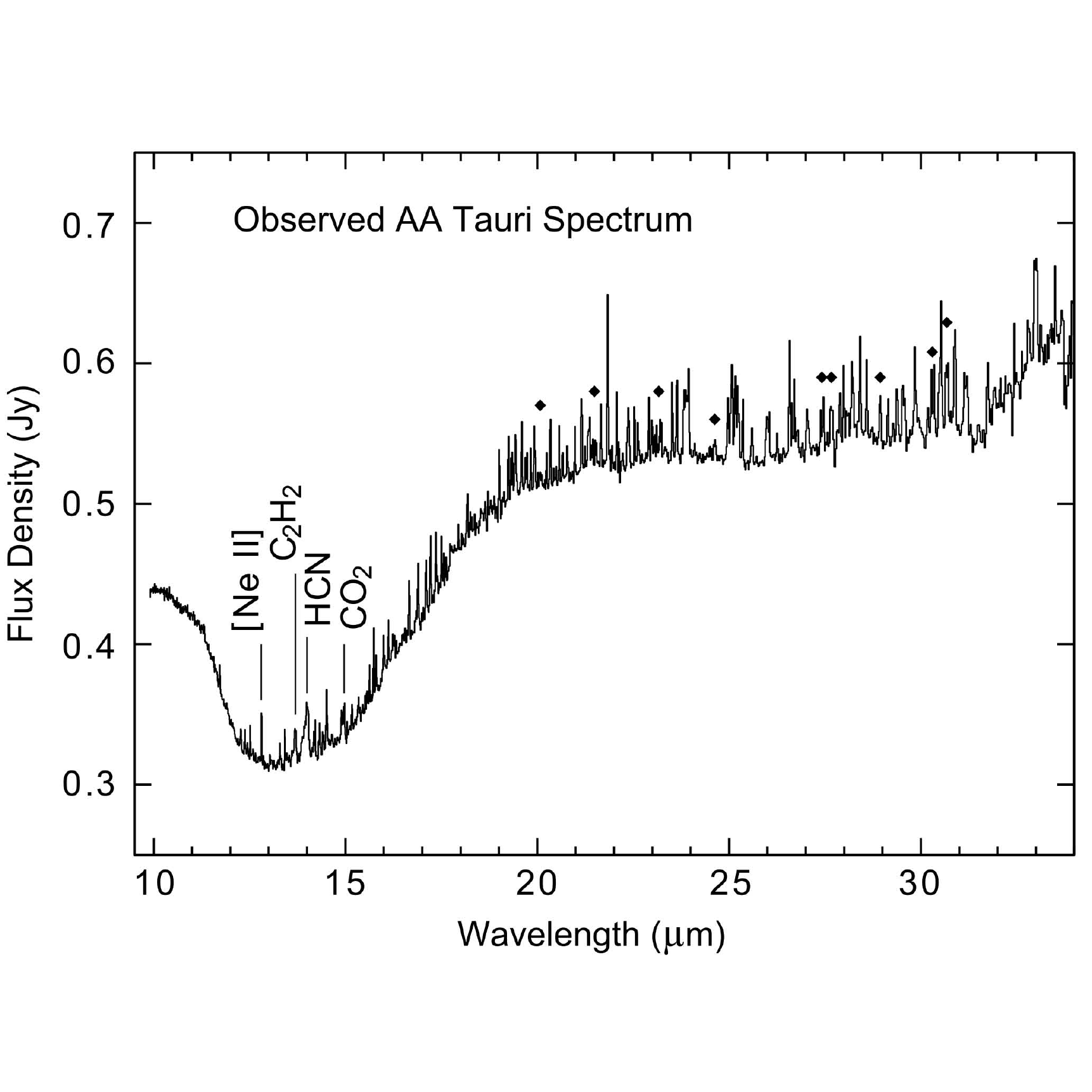}
\caption{Infrared spectrum of the classical T Tauri star AA Tau taken by the Spitzer Space
Telescope.  Vibrational modes of C$_2$H$_2$, HCN, and CO$_2$ along with a transition of  [Ne II] are labelled.  Diamonds mark detections OH rotational transitions.  Numerous rotational transitions of water vapor (not marked) are spread throughout the spectrum.  Figure and discussion published by \citet{cn08}.
\label{fig:cn}}
\end{figure}

\begin{figure}
\includegraphics[width=13.5cm]{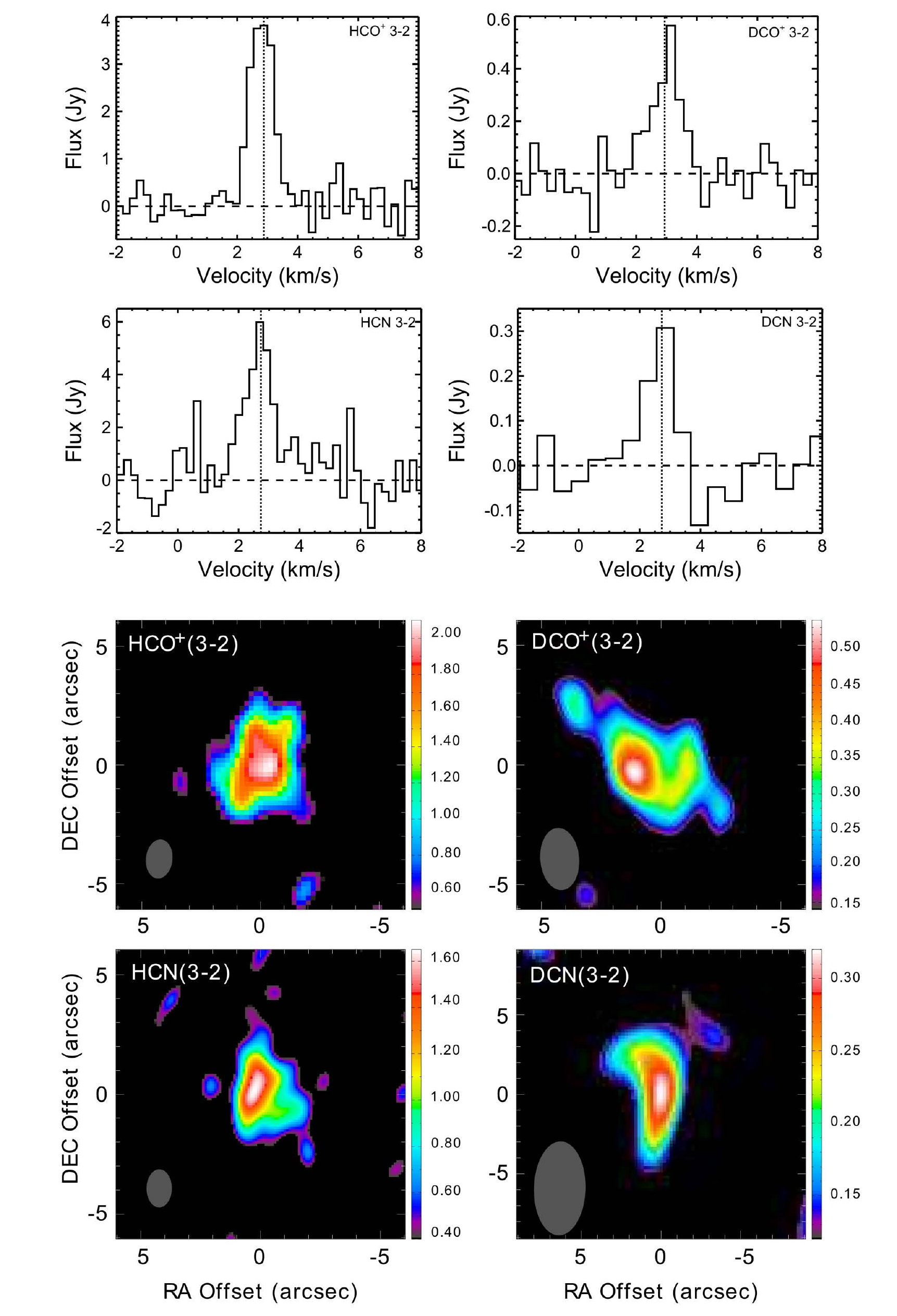}
\caption{Figures showing sample spectra and integrated emission maps of the    $J = 3-2$ transition of  HCO$^+$, DCO$^+$, HCN, and DCN.   Observations from the Submillimeter Array and figure published by \citet{qi_chem}.
\label{fig:qi08}}
\end{figure}

 With these techniques, a simple understanding of the chemistry is emerging.     One underlying result is that the molecular abundances are generally reduced when compared to those measured in the interstellar medium \citep{dgg97, bergin_ppv, dutrey_ppv}.  This is due to the freeze-out of molecules in the dense cold midplane, which will be discussed later in this Chapter.   Closer to the star there exists species specific ``snow-lines'' where molecular ices evaporate.  In this regard,   the presence of hot abundant water in the inner ($< 2$ AU) disk has been confirmed in some systems \citep{cn08, salyk08}.   A sample of these results is shown in Fig.~\ref{fig:cn} where a Spitzer spectrum of a T Tauri star is shown.  
This spectrum reveals not only the presence of water vapor inside the snow-line, but also abundant pre-biotic organics \citep[see also][]{lahuis06}.   It is also becoming possible to resolve chemical structure within protoplanetary systems \citep[e.g.][]{dutrey_ppv}.   A sample of what is currently possible is shown in Fig.~\ref{fig:qi08} where we show the integrated emission distributions of HCO$^+$, DCO$^+$, HCN, and DCN.   As can be seen the emission of deuterium bearing DCO$^+$ is off-center when compared to HCO$^+$, which is indicative of an active deuterium chemistry, at least for that species.
Finally, in the case that most  species containing heavy elements (with dipole moments) are frozen onto grains in the dense midplane then two species, H$_2$D$^+$ and D$_2$H$^+$, will remain 
viable probes \citep{cecc_h2dp}.   Some complex organics have been detected in the mm-wave and a summary of disk organic chemistry is given by \citet{hs08}.

One key factor for modeling disk chemistry is the dissipation timescale of the gas rich disk.
The molecular detections listed in Table~\ref{tab:obs} are predominantly from systems with
ages of $\sim 1 - 3$ Myr,
demonstrating that the gas rich disk clearly
exists for at least a few Myr.   In addition,  
 some gas-rich systems, such as TW Hya, are observed at 10 Myr.
There is growing evidence of increased dust evolution (i.e. grain growth and possible planet formation) in the inner disk (r $< 10 - 30$ AU) of some systems \citep[Calvet, this volume;][]{nsm07}.   Systematic surveys of disk dust emission in stellar clusters has provided some of the strongest limits on the evolutionary timescales for solids.
Based on these studies it is estimated that the timescale for 
 grain growth, settling, and potentially planetesimal formation, is $\sim 1 - 10$ Myr \citep[e.g.][]{haisch_disk, cieza08}.   The disk gas accretion rate decays over similar timescales, although there is a large scatter in the data \citep{muz00}.    Beyond disk accretion, direct observations of  {\em gas} dissipation is limited by our observational capabilities.  However, a Spitzer search for gas emission lines in systems spanning a wide range of ages has suggested that the gas disk dissipates on similar timescales as seen for dust evolution \citep[i.e. $\sim 1 - 10$ Myr;][]{pascucci_feps}.   It is worth noting that the available data, while showing an general decline with age,  exhibits a large degree of scatter.   Thus there are systems that have lost their disk in $\sim 1$ Myr, while others exhibit rich gaseous disks at 10~Myr \citep{cieza07}.

\section{Equilibrium and Kinetic Chemistry}
\label{sec:chem}

In the following we will outline the basic methods and chemical processes that are generally used to examine chemistry in protoplanetary disks.  
In the past, the primary method to explore disk chemistry involved assuming a gas-solid mixture in fixed pressure and temperature with solar composition and determining resulting composition in thermodynamic equilibrium.   This method has been quite successful in explaining various trends observed in meteorites and planetary bodies.   For a summary of these efforts see \citet[][and references therein]{prinn_ppiii, davis_geochem}.   More recently, in part due to observations of an active chemistry on the disk surface, it has been recognized that for much of the disk mass the temperatures and pressures are too low for the gas to reach equilibrium and the gas kinetic chemistry must be considered.   In the following we will describe the basic techniques for both of these cases, with a brief discussion regarding the regimes where each is relevant.

\subsection{Thermodynamic Equilibrium}
\label{sec:te}

Thermodynamic equilibrium is defined as the state of maximum entropy or, the state of minimum Gibb's free energy of the system.   It is the state towards which all chemical systems are being driven if one waited ``forever and a day''.
A key aspect of these calculations is that the final composition depends on the pressure, temperature, and elemental composition, but not the initial chemical state.   For the following the author is grateful to C. Cowley for numerous discussions and providing an outline of the methodology \citep[see][]{cowley}. 

The simplest method involves using the equilibrium constants.
For every reaction an equilibrium can be defined.   
For example in the following reaction:

\begin{equation}
{\rm 2H + O \rightleftharpoons H_{2}O}
\end{equation}

\noindent the ratio of partial pressures ($p_X = n_X k T$) can be determined from the equilibrium constant $K_p$(H$_2$O) which is a tabulated quantity (c.f. JANAF tables from NIST).   The equilibrium constant can be expressed as:

\begin{equation}
K_p({\rm H_2O}) = {\rm \frac{p(H_2O)}{p^2(H)p(O)}} = exp[-\Delta G^{\circ}({\rm H_2O})/RT].
\end{equation}

\noindent Where $R$ is the gas constant and the $\Delta G$ is the Gibbs free energy, which for this reaction is,

\begin{equation}
\Delta G^{\circ}({\rm H_2O}) = \Delta G^{\circ}_f ({\rm H_2O}) - 2\Delta G^{\circ}_f({\rm H}) - 
\Delta G^{\circ}_f({\rm O}).
\end{equation}

\noindent The $o$ superscript denotes that the substance is in its standard state given the temperature (i.e. liquid, gas, solid) and  subscript $f$ refers to ``of formation''.   In a system with multiple species at fixed pressure and temperature we can use the equilibrium constants to determine the composition.
 For an example, lets take the following system:

\begin{eqnarray}
{\rm CO} & \rightleftharpoons & {\rm C + O}\\
 {\rm CN} & \rightleftharpoons &   {\rm C + N}.
\end{eqnarray}

\noindent We know the following:

\begin{eqnarray}
K_{{\rm CO}}(T) &=& \frac{p_{\rm C} p_{\rm O}}{p_{\rm CO}},\\
K_{\rm CN}(T) &= &\frac{p_{\rm C} p_{\rm N}}{p_{\rm CN}}.
\end{eqnarray}

\noindent We can now define ``fictitious pressures'', $P_{\rm C}$, $P_{\rm O}$, and $P_{\rm N}$ which can be expressed,

\begin{eqnarray}
P_{\rm O} &=& p_{\rm O} + p_{\rm CO} = p_{\rm O} + \frac{p_{\rm C} p_{\rm O}}{K_{\rm CO}}\\
P_{\rm N} &=& p_{\rm N} + p_{\rm CN} = p_{\rm N} + \frac{p_{\rm N} p_{\rm C}}{K_{\rm CN}}\\
P_{\rm C} &=& p_{\rm C} + p_{\rm CN} + p_{\rm CO} = p_{\rm C} + \frac{p_{\rm C} p_{\rm N}}{K_{\rm CN}} + \frac{p_{\rm C} p_{\rm O}}{K_{\rm CO}}
\end{eqnarray}

\noindent and the pressure (mass) balance equation,

\begin{equation}
P = p_{\rm C} + p_{\rm N} + p_{\rm O} + p_{\rm CO} + p_{\rm CN}
\end{equation}

\noindent From these you can solve for the partial pressures and composition.

 There are some other more commonly used methods in the literature, which we will outline below.   The Gibb's free energy ($G$) is defined by:

\begin{equation}
G = H -TS = E + PV - TS
\end{equation}

\noindent where $H$ is the enthalpy, $E$ the energy,  $S$ the entropy.  $P$, $T$, and $V$, are, respectively, the pressure, temperature, and volume.  From this the change in energy is,
 
\begin{equation}
dG = dE + PdV  + VdP - TdS - SdT,
\end{equation}

\noindent which, using the 1st law of thermodynamics, reduces to,

\begin{equation}
dG = VdP - SdT.
\end{equation}

\noindent If this is an ideal gas than $PV = nRT$, and

\begin{equation} 
(dG)_T = VdP = \frac{nRT}{P}dP,
\end{equation}
 
\begin{equation}
G_B - G_A = \int^{p_B}_{p_A}\frac{nRT}{P}dP = nRTln\left(\frac{p_B}{p_A}\right)
 \end{equation}

\noindent In the case of departures from an ideal gas, then:

\begin{equation}
 G_B - G_A = \int^{p_B}_{p_A}\left(V - \frac{nRT}{P}\right)dP + nRTln\left(\frac{p_B}{p_A}\right) = nRTln\left(\frac{f_B}{f_A}\right)
 \end{equation}
 
 \noindent where $f$ is the fugacity, which is essentially a pressure term for a non-ideal (real) gas.   It is equivalent to the partial pressure if the gas is ideal.
 It is now common to re-write this in terms of some known or measured state (e.g. 1 atm where $f = f_0$):
 
 \begin{equation}
 G_B = G^0 + nRTln\left(\frac{f_B}{f_0}\right) = G^0 + nRTln(a)
 \end{equation}

\noindent where $a$ is the activity which is defined as $f/f_0$.  
 If we have an ensemble of atoms and molecules we can write out the total Gibbs free energy:
 
 \begin{equation}
G = \Sigma n_{i,p}(\Delta G_{i,p} + RTln\,a_{i,p})
 \end{equation}
 
 \noindent where $n$ is the number of moles of species $i$ in phase $p$ (either pure condensed, condensed, or gaseous).  The final equation is for mass balance normalized to the assumed initial composition.   Given these sets of equation the minimized value of the Gibb's free energy determines the equilibrium composition.    Thermodynamic data are available via the JANAF tables \citep{chase98}, although the tables are not complete.  Some useful references that more outline this procedure and  more directly discuss issues regarding thermodynamic data are \citet{burrows99} and \citet{lodders02}.

 \subsubsection{Condensation}
\label{sec:conden}

The condensation temperature of an element as defined by \citet{gpp} is ``the temperature at which the most refractory solid phase containing that element becomes stable relative to a gas of solar composition''. 
 A full discussion of this topic is beyond the scope of this chapter and the reader is referred to \citet{1972GeCoA..36..597G}, \citet{gpp},  \citet{2003ApJ...591.1220L} (and references therein). 
  However, as discussed in \S~\ref{sec:obs}, these calculations lie close to the heart of understanding the composition of some meteorites  and as such warrant a brief discussion.  The essential idea is that as the nebula gas cools (at constant pressure) refractory minerals condense from the gas and this sequence depends on the condensation temperature of the various potential refractory minerals.  When a given element condenses one must pay attention to mass balance as some (or all) of its abundance is now unavailable for the gas in equilibrium.   As an example, Corundum (Al$_2$O$_3$) is one the first minerals to condense.  This is consistent with the detection of this minerals in Ca-Al rich inclusions in chondrites and the fact that Ca-Al inclusions provide the oldest age measured in the Solar System \citep{amelin02}. 
   An excellent further discussion of this topic is found in the references above and also in \citet{davis_messii}.    

\subsubsection{Fischer-Tropsch Catalysis}

Fischer-Tropsch catalysis is the name for a subset of high temperature reactions wherein
CO (and CO$_2$) with H$_2$ is converted to methane and other hydrocarbons on the surfaces of transition metals, such as Fe and Ni.   If Fe/Ni grains are present then Fischer-Tropsch catalysis is an effective method to create hydrocarbons from gas-phase CO at temperatures where the reaction is inhibited due to high activation energies ($T <$ 2000~K).    The conversion of CO to organics via this process has been explored by numerous authors and the reader is referred to \citet{fp89}, \citet{kt01} and \citet{sekine05}, and references therein.   The efficiency of Fischer-Tropsch catalysis is highly dependent on the gas temperature and pressure.   The efficiency, measured at pressures relevant to giant planet subnebula\footnote{Giant planet subnebula refers to the collapsing pocket of the surrounding disk gas out of which the giant planet forms.   Both the gas/ice giant and its satellites form within the subnebula. } (0.09--0.53 bar) peaks at $\sim$550 K with reduced efficiency at either lower and higher temperatures \citep{sekine05}, but see also \citet{lc00} for measurements at lower pressures.   At higher temperatures the surface of the catalyst is poisoned by the conversion of surface carbide to unreactive graphite.  At lower temperatures the efficiency will ultimately be limited by the incorporation of Fe into silicates, the formation of Fe$_3$O$_4$ coatings \citep{pf89}, or coatings by organic and water ice \citep{sekine05}.     It is also worth noting that ammonia could also be formed from N$_2$ via a similar process called Haber-Bosch catalysis.

\subsubsection{Clathrate Hydrates}

\citet{ls85} provided a detailed analysis of the thermodynamics and kinetics of clathrate hydrates at relevant conditions for the solar nebula.  A clathrate hydrate is a water ice crystal in which the lattice is composed of water molecules with open cavities that are stabilized by the inclusion of small molecules of other chemical species.  The inclusion of additional molecules in the lattice is actually key to the stability of the clathrate \citep{buffett00, devlin01}.  Because this support the clathrate latices are more open then either the regular structured lattice of crystalline ice or irregular structured lattice of amorphous ice \citep[see, e.g.][]{tanaka06}.

 The relevance of these compounds for nebula chemistry is that they provide a means of trapping volatile species (noble gases, CO, N$_2$) into ices at temperatures well above their nominal sublimation temperature.    Thus volatiles can be provided to forming planets to be later observed in their atmospheres.   \citet{ls85} suggest that clathrate hydrates can be important, but only if fresh ice is exposed to the surrounding gas via frequent collisions or perhaps if ices evaporate and condense upon an accretion shock \citep[e.g.,][]{lunine91}.   

\subsubsection{Kinetic Inhibition}

It was recognized that the conditions in the solar nebula may not always be in thermodynamic equilibrium.     In particular, the timescale of nebular evolution (e.g. the gas lifetime) could be shorter than reaction timescales and therefore equilibrium would not be attained.   Alternately, the dynamical mixing of  gases from colder regions within the nebula to the warmer central regions could preclude the formation of certain species which would otherwise be favored given local conditions.   Thus knowledge of the reaction kinetics is required.  Cases where kinetics limit the outcome of thermodynamic equilibrium are labeled as kinetic inhibition by \citet{lp80}.  It is also referred to as quenching in the literature.   As an example, 
in equilibrium CH$_4$ should dominate over CO in the inner solar nebula and also within giant planet subnebula.   However, the chemical timescale for CO conversion is $\sim 10^9$ s at  $T_{gas}$ $\sim$ 1000~K and longer at lower temperatures \citep{lp80}.  Estimates suggest that dynamical timescales are shorter,  and the zone where CH$_4$ dominates therefore shrinks considerably  \citep[e.g.][]{fp89, ma06}.     In general, these models derive the equilibrium results and then explore whether kinetics would change the predicted results \citep[see,][for a discussion]{smith98}.


 \subsection{Chemical Kinetics}
 
 For much of the disk mass the density and temperature are low enough that the gas does not have sufficient time to reach equilibrium and one needs to explore the time-dependence of the chemistry.   The exact temperature and density (and various combinations) where equilibrium calculations are not valid is presently unclear, and we include a brief discussion of this in \S~\ref{sec:valid}.  However, it is certain that  within regions of the disk where the gas temperature is below 100 K (which is most of the disk mass) that the chemical kinetics should be considered.
 
   Fortunately, observations (see \S~\ref{sec:obs}) appear to suggest that the observed chemistry on tens of AU scales is similar to that seen in the dense ($n > 10^6$ \cc ) regions of the interstellar medium exposed to energetic radiation fields.   Models developed for application in this regime
 \citep{herbst_arpc} have been readily applied to protoplanetary disks \citep[see, e.g.][]{aikawa97, bauer97, aikawa99}.   A key feature of the interstellar models is that the cold ($T_{gas} < 100$ K) temperatures require exothermic reactions, while the low ($n \sim 10^3 - 10^6$ \cc ) density limits reactions to two bodies.   At these energies typical reactions between two neutrals are endothermic or have an activation barrier.  Therefore, exothermic reactions between ions and neutrals are believed to dominate \citep{hk73, watson76}.    However, given the strong radial gradients in density and temperature, it is certain that in some instances reactions with barriers are important and 3-body reactions can also be activated \citep{willacy98, tg07}.   
 
 In the disk context the temperature is often low enough that water and other more volatile species will freeze (or even be created in situ) on grain surfaces. 
Thus the inclusion gas-grain interaction is a key piece of any successful disk chemistry model.   Below we will first describe the basic gas-phase chemistry and how networks are created and solved.  Then we extend the discussion to include grain chemistry.

\subsubsection{Gas-Phase Chemistry}
\label{sec:gas}

  In Table~\ref{tab:reactions} we provide a listing of common chemical reaction types and typical rates.   To be active the gas-phase chemistry requires a source of ionization, which in the disk can either be cosmic rays, X-rays, or active radionuclides  (\S~\ref{sec:ion}).   The ionization of H$_2$  ultimately produces \Hthreep , which powers the ion-molecule chemistry.  A general feature of this chemistry is the subsequent creation of complex molecular ions which undergo dissociative recombination to create both simple and complex molecules.   To illustrate the various features of the chemistry we will examine the chemistry of water which has significance in the disk context.

\begin{threeparttable}
\small
\caption{Typical Reaction Types and Coefficients}
\begin{tabular}{lll}
\hline
\multicolumn{1}{c}{Reaction Type} &
\multicolumn{1}{c}{Example} &
\multicolumn{1}{c}{Typical Rate Coefficient} \\\hline\hline\\
Ion-Neutral & H$_3^+ +$ CO$ \rightarrow$ HCO$^+ +$ H$_2$ & $\sim 10^{-9}$ cm$^3$ s$^{-1}$ \\\\
Radiative Association\tnote{a} & C$^+ +$ H$_2 \rightarrow$ CH$_2^+ + h\nu $ & $\sim 10^{-17}$ cm$^3$ s$^{-1}$\\\\
Neutral Exchange & O $+$ OH $ \rightarrow$ O$_2 +$ H & $\sim 10^{-11} - 10^{-10}$ cm$^3$ s$^{-1}$ \\\\
Charge Transfer & C$^+ +$ S $\rightarrow$ S$^+ +$ C & $\sim 10^{-9}$ cm$^3$ s$^{-1}$ \\\\ 
Radiative Recombination\tnote{a} & C$^+ + e^- \rightarrow$ C $+ h\nu$ & $\sim 10^{-12}$ cm$^3$ s$^{-1}$ \\\\
Dissociative Recombination\tnote{a}$\;$ & H$_3$O$^+ + e^- \rightarrow$ H$_2$O $+$ H & $\sim 10^{-6}$ cm$^3$ s$^{-1}$ \\\\
Photoionization & C $ + h\nu \rightarrow$ C$^+$ $+$ e$^-$& $\sim 10^{-9}e^{-\gamma A_v}$ cm$^3$ s$^{-1}$ \\\\
Photodissociation & H$_2$O $+ h\nu \rightarrow$ O $+$ OH & $\sim 10^{-9}e^{-\gamma A_v}$ cm$^3$ s$^{-1}$ \\\hline
\end{tabular}
\begin{tablenotes} \footnotesize
\item [a]  Sample rate coefficients derived at a gas temperature of 40 K using the UMIST database \citep{woodall07}.\\
\item [b] The depth dependence of photoionization rates and photodissociation rates is often parameterized by e$^{-\gamma A_v}$.  In this form $\gamma$ includes factors such as the shape and strength of the radiation field, the absorption and scattering due to dust, and the wavelength dependence of the photoionization or photodissociation cross-sections.   For greater discussion please see \citet{vd88}, \citet{roberge91}, and \citet{vd06}. \\
\end{tablenotes}
\label{tab:reactions}
\end{threeparttable}

The oxygen chemistry begins with the 
ionization of \Htwo\ and the formation of the trihydrogen ion
(\Hthreep ).  \Hthreep\ will react with O and a sequence of rapid
reactions produces \HthreeOp . 
The next step involves the dissociative recombination of \HthreeOp\ 
with electrons:

\begin{eqnarray}
\rm{H_3O^+ + e^-}  &\rightarrow & \rm{H + H_2O}\;\;\;\;\;\;\;\;\;\;\: f_1\\
& \rightarrow & \rm{OH + H_2}\;\;\;\;\;\;\;\;\;\;\:  f_2 \nonumber\\
&\rightarrow & \rm{OH + 2H}\;\;\;\;\;\;\;\;\;\;\: f_3 \nonumber\\
&\rightarrow & \rm{O + H + H_2}\;\;\;\;\;\;\:f_4\nonumber
\end{eqnarray} 

\noindent where $f_{1-4}$ are the branching ratios.\footnote{
Two measurements using a storage ring give roughly consistent results
\citep{jensen_h3op, vc_h3op}.   We provide here the latest measurement:
$f_1 = 0.25$, $f_2 = 0.14$, $f_3 = 0.60$, and
$f_4 = 0.01$.   Using a different technique (flowing afterglow), \citet{williams_h3op} find 
$f_1 = 0.05$, $f_2 = 0.36$, $f_3 = 0.29$, $f_4 = 0.30$.  }  Via this sequence of reactions water vapor is then created.    The primary {\em gas-phase} destruction pathway in shielded gas is via a reaction with He$^+$ (another ionization product); at the disk surface photodissociation also plays a key role.    It should be stated that the example above provides only the primary formation and destruction pathways under typical conditions (i.e. where ionizing agents are available).  In addition, water participates in numerous other reactions both as reactant and product and some water can be created via other pathways, albeit at reduced levels \citep[e.g.][]{bergin_impact}.

At temperatures below a few hundred Kelvins  ion-neutral chemistry dominates.  Above this temperature, two neutral-neutral reactions rapidly transform all elemental oxygen in the molecular gas into water vapor
\citep{edj78, wg87}:

\begin{eqnarray}
\rm{O  +   H_2} &   \rightarrow & \rm{  OH   +   H}\;(E_a= 3160\;\rm{K}),\\
\rm{OH + H_2} & \rightarrow & \rm{H_2O + H}\;(E_a= 1660\;\rm{K}).
\end{eqnarray}

\noindent This mechanism (along with evaporation of water ice) will keep water vapor the dominant oxygen component in the gas within a few AU of the forming star.

To model these reactions one creates a series of ordinary differential equations, one for each species included, accounting for both formation rates and destruction rates: $dn/dt = formation - destruction$.  For water
(excluding the numerous reactions not mentioned above) the equation is as follows:

\begin{eqnarray}
\frac{dn(\rm{H_2O})}{dt} &=& n(\rm{H_3O^+})n(\rm{e^-})\alpha_r 
  + n(\rm{OH})n(\rm{H_2})k_1 \nonumber\\
  &-& n(\rm{H_2O})[n(\rm{He^+})k_2 + k_{\gamma, A_v = 0}e^{-1.7A_v}] 
\label{eq:rate}
\end{eqnarray}

\noindent In the equation above $\alpha_r$ is the recombination rate of \HthreeOp\ \citep{jensen_h3op} and $ k_{\gamma, A_v = 0}$ is the unshielded photodissociation rate.   Values for this rate and others can be found in either of the two available astrochemical databases \citep[one originally maintained by E. Herbst at Ohio State or the other at UMIST:][]{umist06}.
These rate networks have thousands of reactions occurring over a wide-range of timescales  and thus it is a stiff system of ordinary differential equation (ODE).   A number of ODE solvers have been developed to solve these systems with the output of composition as a function of time.  A summary of some of the available codes for astrochemical applications is given by \citet{nejad05}.

\subsubsection{Gas-Grain Chemistry} 
\label{sec:gas-grain}

In the dense ($n > 10^5$ \cc ), cold ($T < 20$ K) centers of condensed molecular cores, which are the stellar birth sides, there is ample evidence for the freeze-out of many neutral species onto grain surfaces \citep{bt_araa}.   This even includes the volatile CO molecule that only freezes onto grains when temperatures are below 20 K \citep{collings_cobind}, as opposed to water which freezes at 110 K for interstellar pressures \citep{fraser_h2obind}.    The outer disk midplane ($r  > 20-40$ AU) is believed to be similarly cold  with densities that are several orders of magnitude higher than in pre-stellar cores where volatile CO freeze-out is readily detected.  
%
 Thus to model disk kinetic chemistry it is necessary to model both the deposition of species onto grain surfaces in tandem the mechanisms which desorb frozen ices.   A number of such desorption mechanisms have been identified in the literature including thermal evaporation, X-ray desorption, and UV photo-desorption.     Each of these will be described below and are generally included in the rate equations (Eqn.~\ref{eq:rate})  as source or loss terms with rate coefficients given below.

\begin{figure}[t]
\includegraphics[width=11.5cm]{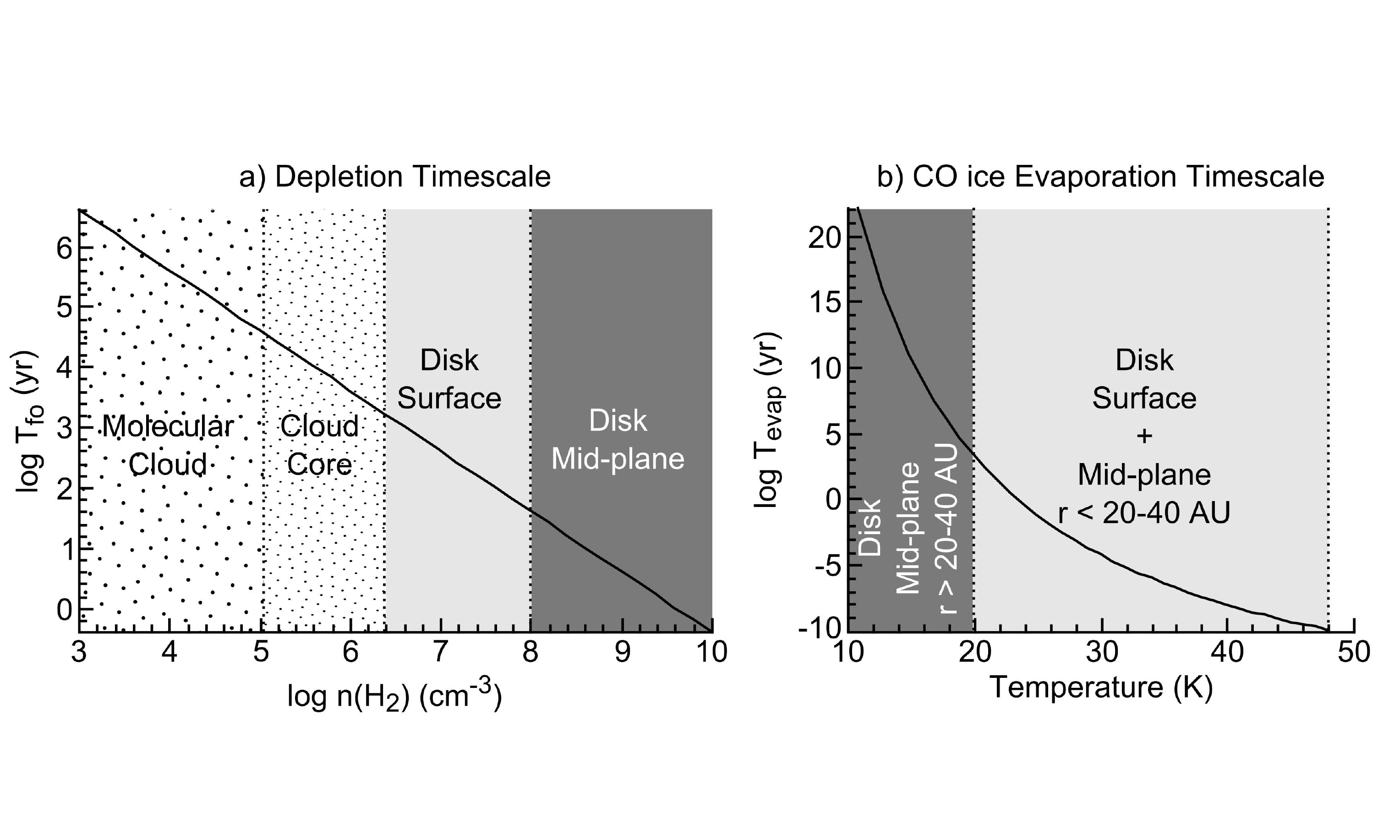}
\caption{{(a)}  CO freeze-out timescale as a function of density of molecular hydrogen.  (b) Thermal evaporation timescale for CO molecules as a function of dust temperature.
\label{fig:time}}
\end{figure}

\medskip
\noindent\underline{Gas-phase Freeze-out:}
Atoms and molecules freeze onto grain surfaces or adsorb with a 
rate of $k_{fo} = n_{g}\sigma_g v S$ s$^{-1}$.   Here $\sigma_g$ is the grain cross-section, $v$ is the mean velocity of the Maxwellian distribution of gaseous particles, $n_g$ the space density of grains, and $S$ is the sticking coefficient (i.e. how often a species will remain on the grain upon impact).  
Sticking coefficients for molecular species are generally thought to be unity at low $T$ which is supported by theoretical \citep{burke_hollenbach} and laboratory work \citep{bisschop_stick}.
For an interstellar grain size distribution\footnote{In the interstellar medium grains are inferred to exist with a size distribution that follows a power-law of $dn(a)/da \propto a^{-3.5}$, with $a$ as the grain radius \citep{mrn}.   }
 with a lower cutoff of 20~\AA , $n_g \sigma_g \simeq 2 \times 10^{-21}n$ cm$^{-1}$ \citep[see discussion in][]{hkbm08}.   In disk systems the process of grain coagulation and settling will locally reduce the grain surface area and this value is only a starting point for potential situations.   Assuming this surface area, the freeze-out timescale, $\tau_{\,fo}$, is:
 
\begin{equation}
\tau_{\,fo} = 5 \times 10^3 {\rm yrs} \left(\frac{10^5 cm^{-3}}{n}\right) \left(\frac{20 K}{T_{gas}}\right)^{\frac{1}{2}}\sqrt{A}
\end{equation}

\noindent where $A$ is the molecule mass in atomic mass units.   Fig.~\ref{fig:time} demonstrates that within the disk the freeze-out time for CO is on the order of tens of years or below.  Thus, without effective means to remove molecules from the grain surface, most heavy molecules freeze onto grain surfaces. In the astronomical literature this timescale is often referred to as the depletion timescale.  Because molecular abundances can ``deplete'' by freeze-out or gas-phase destruction, we adopt here the more descriptive terminology.

\medskip
\noindent\underline{Grain Binding Energies:}

The removal of ices from the refractory grain core requires overcoming the strength of the bond to the surface.   Because low temperatures preclude the breaking of chemical bonds, molecules are not expected to be chemically bound to grain surfaces \citep[although chemical bonds may be important at higher temperatures;][]{ct04}.
Instead, the 
approaching molecule has an induced dipole interaction with the
grain surface or mantle and are bound through weaker
Van der Waals-London interactions called physical adsorption
\citep{kittel}.  Van der Waals interactions are proportional to the product
of the polarizabilities ($\alpha$) of the molecule and the nearby surface
species ($E_b \propto \alpha_{mol}\alpha_{surface}$).  Because of this property
binding energies on a silicate surface have been calculated
using the extensive literature on the physical adsorption of species onto
graphite.  Accounting for differences in polarizability these can
be scaled to silicates \citep{ar77}.
\citet{hhl92}  re-examined this issue and provide a
sample list of estimated values that have been widely used.
These estimates need to be revised for cases where actual experimental information exists. 
Measurements have been now been performed for a number  of key species, with a listing given in  Table~\ref{tab:binde}.

\begin{center}
\begin{table}[t]
\caption{Binding Energies of Important Ices$^a$}
\centering
\begin{tabular}{lll} 
\hline
\multicolumn{1}{c}{Species} &
\multicolumn{1}{c}{$E_b$/k} &
\multicolumn{1}{c}{Reference} \\\hline\hline
\HtwoO\ & 5800 & \citet{fraser_h2obind} \\
CO & 850 & \citet{collings_cobind} \\
N$_2$ & 800 & \citet{bisschop_stick}\\
\NHthree\ & 2800 & \citet{bb07}\\
\CHthreeOH\ & 5000 & \citet{bb07}\\\hline
\multicolumn{3}{l}{$^a$ \footnotesize Values are from laboratory work only}\\
\multicolumn{3}{l}{\footnotesize using pure ices.}
\end{tabular}
\label{tab:binde}
\end{table} 
\end{center}

\medskip
 \noindent\underline{Thermal Evaporation:}

 The rate of evaporation from a grain with a temperature of $T_{dust}$, is given by the Polanyi-Wigner relation \citep{tielens_book}:  

\begin{equation}
k_{evap,i} = \nu_{0,i} exp(-E_i/kT_{dust})\;\;(s^{-1}).
\end{equation}

\noindent $\nu_0$ is the vibrational frequency of molecule $i$ in the potential well,

\begin{equation}
\nu_i = 1.6 \times 10^{11} \sqrt{(E_i/k)(m_H/m_i)}\;{\rm s}^{-1},
\label{eq:nugr}
\end{equation}

\noindent with $E_{b,i}$  the binding energy of species $i$.  Following \citet{hkbm08} the sublimation or freeze-out temperature can be derived by setting the flux of desorbing molecules ($F_{td,i}$) equal to the flux of species adsorbing from the gas.    Thus,

\begin{equation}
F_{td,i} \equiv N_{s,i} k_{evap,i} = 0.25n_i v_i,
\end{equation}

\noindent where N$_{s,i}$ is the number of adsorption sites per cm$^2$
($N_{s,i} \sim 10^{15}$ cm$^{-2}$), $n_i$ and $v_i$ are the gas-phase number density and thermal velocity of species $i$.    Solving for the dust temperature we can derive the sublimation temperature for species $i$ \citep{hkbm08}:

\small
\begin{equation}
T_{sub,i} \simeq \left(\frac{E_{b,i}}{k}\right)\left[57 + ln\left[\left(\frac{N_{s,i}}{10^{15} {\rm cm^{-2}}}\right) \left(\frac{\nu_i}{10^{13}\;{\rm s^{-1}}}\right) \left(\frac{1\;{\rm cm^{-3}}}{n_i}\right) \left(\frac{10^4\;{\rm cm\; s^{-1}}}{v_i}\right) \right] \right]^{-1}
\label{eq:tsub}
\end{equation}
\normalsize 

\noindent The above equation assumes that there is at least one monolayer of the particular species present on the grain surface.  In the case where only a partial monolayer exists then the thermal desorption rate per unit grain surface area is $f_{s,i}N_{s,i}k_{evap,i}$, where $f_{s,i}$ is the fraction of the $N_{s}\pi a^2$ surface sites occupied by species $i$.
In Fig.~\ref{fig:time}(b) we show a plot of the CO evaporation timescale as a function of grain temperature.  Comparing to the freeze-out timescale in Fig.~\ref{fig:time}(a) suggests that CO will be frozen in the form of ice throughout much of the outer ($r > 20-40$ AU) disk midplane.  However, on the warmer disk surface (see Fig.~\ref{fig:overview}) CO will remain in the gas phase.  This is consistent with observations \citep{vanz_etal01}.    

\begin{figure}[t]
\includegraphics[width=11.5cm]{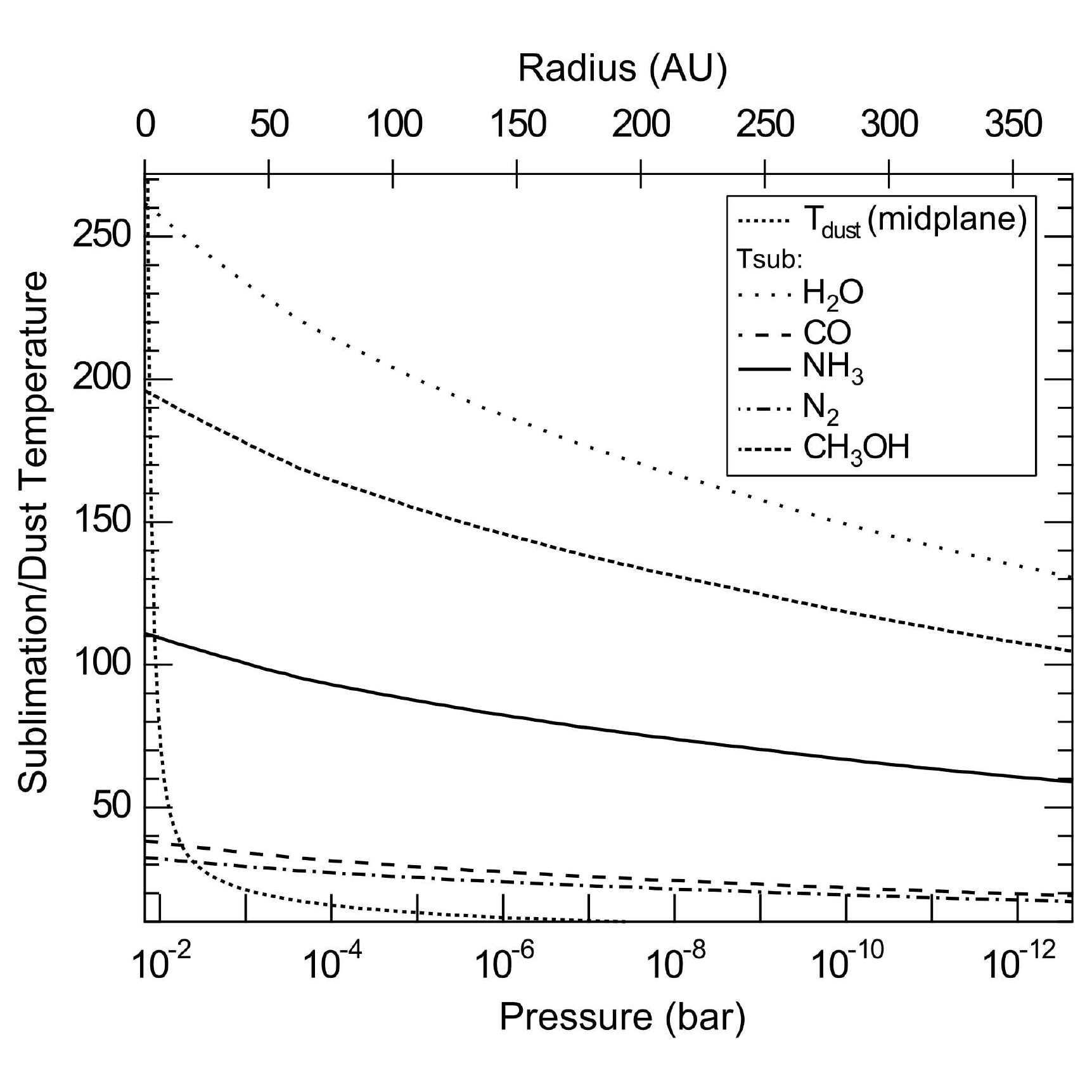}
\caption{Plot of the sublimation temperature for various molecular ices, and the midplane dust temperature,  as a function of disk pressure and radius for a disk model taken from \citet{dch01}.
\label{fig:tsub}}
\end{figure}

Another interesting aspect of Eqn.~\ref{eq:tsub} is the dependence on gas density.  In Fig.~\ref{fig:tsub} we provide a plot of the sublimation temperature (and dust temperature) as a function of disk pressure and radius for a model solar analog disk \citep[taken from][]{dch01}.   This calculation uses the binding energies given in Table~\ref{tab:binde}.    Thus most volatile ices are frozen beyond  tens of AU with various species returning to the gas in a fashion that depends on their relative binding strength to the grain surface.   It should also be noted that experiments suggest that the return of species to the gas is not entirely as simple as described above.   The evaporation also depends on the composition of the mantle and the relative disposition of its components \citep{viti_evap}.

\medskip
\noindent\underline{Photodesorption:}

The direct absorption of an ultraviolet (UV) photon by a molecule adsorbed on the grain surface puts an electron in an excited
state.  If the interaction between the excited state
and the binding surface is repulsive, then the excited molecule may be
ejected from the surface \citep{watson76}.  The rate of photodesorption per adsorbed molecule is given by \citep{boland_dejong, hkbm08}:

\begin{equation}
R_{pd} = n_{gr} \pi a_d^2 \epsilon Y G_0 F_0 exp(-1.8A_{v})\; {\rm cm^{-3}s^{-1}}.
\end{equation}

\noindent In this equation  $Y$ is the photodesorption yield (number of molecules
desorbed per incident photon), $\epsilon$ is the fraction of species that are found in the top few monolayers, $a_d$ is the grain radius, and $G_0$ is the UV
field enhancement factor in units of the mean interstellar UV field of $F_0 = 10^{8}$ photons cm$^{-2}$ s$^{-1}$
\citep{habing68}.   The stellar UV field and its relation the interstellar radiation field is discussed in \S~\ref{sec:uv}.
For most molecules the yield is unknown and is assumed to be $\sim 10^{-5} - 10^{-3}$ \citep{blg95}.  More recently, there have been measurements for CO \citep[$Y = 3 \times 10^{-3}$ mol/photon;][]{oberg07} and H$_2$O \citep[$Y = T/100 \times  1.3 \times 10^{-3}$ mol/photon;][]{oberg09}.

\medskip
\noindent\underline{X-ray Desorption:}

Young stars are known to be prolific emitters of X-ray photons with emission orders of magnitude above typical levels for main sequence stars \citep{fm99}.  
This emission is capable of penetrating much deeper into the disk atmosphere than UV photons and hence can potentially provide energy to release molecular ices from grains on the disk surface.
 For this chapter we will focus on the effects of these X-rays on the chemistry and refer the reader to the chapter by Feigelson for a discussion of the origin and characteristics of X-ray emission.   X-ray photons are absorbed by the atoms that compose the grain refractory core and mantle with a cross-section that depends on species and wavelength \citep{mm83, henke93}.  The absorption of an X-ray produces  a hot primary electron that generates a number of secondary electrons ($s$) via a process known as the 
 Auger effect.  Depending on the energy of these electrons, and the grain size, they then deposit all or a fraction of their total energy to the grain \citep{ds96}.  This heat diffuses through the grain lattice based on the thermal physics of the material.   Fortunately experiments have demonstrated that the thermal conductivity and specific heats of amorphous solids tend have near universal values to within an order of magnitude over  3 decades in temperature \citep{pohl98}.  This is encouraging for interstellar studies as we can more reliably adopt laboratory estimates for studies of grain heating \citep[see appendix in][]{najita_xray}.

\citet{leger85} explored grain heating by X-rays and cosmic rays adopting the specific heat measurements of amorphous silica from \citet{zp71} in the following form:

\begin{eqnarray}
\rho C_V(T)& = & 1.4 \times 10^{-4}T^2\;J\;cm^{-3}\;K^{-1}\;\; (10 < T < 50\;K)\\ 
\rho C_V(T)& = & 2.2 \times 10^{-3}T^{1.3}\;J\;cm^{-3}\;K^{-1}\;\; (50 < T < 150\;K)
\end{eqnarray} 

\noindent Using this expression the energy increase due to an energy impulse  raises the temperature by (using the specific heat appropriate for $T < 50$ K):

\[
\Delta E(eV) = 3.4 \times 10^4 \left(\frac{a}{0.1\mu m}\right)^3 \left(\frac{T_f}{30\;K}\right),
\]

\noindent where T$_f$ is the final grain temperature and $a$ is the grain size \citep{leger85}.  \citet{najita_xray} explored the question of X-ray desorption in protoplanetary disks using the results of \citet{ds96} and \citet{leger85}.  As one example they found that a 1 keV photon will deposit $\sim 700$eV in a 0.1 $\mu$m grain.  The warm grain cools via evaporation and thermal radiation.   This energy impulse is insufficient to raise a cold ($T_{dust} \sim$ 10 K) grain to above the threshold for desorption of the volatile CO molecule.  However, smaller grains ($a < 0.03 \mu$m) have reduced volume for heat diffusion and equilibrate to temperatures that evaporate CO molecules.
Thus X-ray desorption from the whole grain is not feasible, as  grains with sizes $> 0.1$ $\mu$m are present in the ISM and in disks.   A more efficient method for desorption is found when one considers the fact that grains are likely porous fractal aggregates composed of a number of small grains with a range of sizes.  In this case the absorption of an X-ray by a portion of the grain is capable of local evaporation prior to the heat diffusing throughout the aggregate \citep{najita_xray}.   This is efficient for desorbing the most volatile ices (e.g. CO, N$_2$) on the disk surface, but not water ice.  There is some evidence that this process is active \citep{greaves05}; although the situation is complex because one additional effect of X-ray ionization is the excitation of H$_2$ which generates a local UV field that could release molecules via photodesorption.

\medskip
\noindent\underline{Cosmic ray Desorption:}

Similar to X-rays, cosmic rays absorbed by dust grains can provide non-thermal energy to desorb molecules frozen on grain surfaces.   Cosmic rays also have greater penetrating power, when compared to X-rays and UV, and hence can have a greater impact on the chemistry in well-shielded regions.   The penetration of cosmic rays to the inner tens of AU of the disk is highly uncertain; however, it is likely that cosmic rays can affect the chemistry of the outer disk.  This issue will be discussed later in \S~\ref{sec:ion}, and here we will focus on the effects of non-thermal desorption via cosmic ray impact.    Cosmic rays have energies between 0.02 -- 4 GeV per nucleon, but the energy deposition goes as $Z^2$ such that the less abundant heavier cosmic rays have greater impact on the grain thermal  cycling and ice evaporation \citep{leger85}.   \citet{bringa} noted that the impact of the cosmic ray on a grain is not exactly a question of thermal diffusion (as normally treated), but rather one of a pressure pulse and a melt.   
In general, the effect of cosmic rays is to efficiently remove all ices (including water) from small ($0.01\;\mu$m) grains due to large thermal impulses.
Cosmic ray impacts can also release ices from larger grains via local spot heating.  This is thought to be capable to remove the most volatile ices (CO, N$_2$), but not water ice
\citep{leger85, ch06}.
  A number of groups have looked at this process and there is disparity in the calculations.   A summary of the  various approximations is provided by \citet{rrvw07}.

\medskip
\noindent\underline{Explosive Desorption:}

Experiments of ices irradiated in the lab suggest that when an ice mantle formed at $\sim 10$~K  is exposed to radiation and then warmed to $\sim 25$~K , the rapid movement (\S~\ref{sec:surchem}) and reactivity of radicals leads to an explosion of chemical energy and desorption \citep{sg91}.    There are two issues with this mechanism that are not insurmountable, but make it more difficult to readily place into models.  
(1) The generation of radicals on the surface needs the ice-coated grain to be exposed to the UV radiation field.   A criticism of this is that the UV flux in the lab 
was significantly greater than typical exposure in the well shielded ISM where ice mantles form.
   However,  the strength of the UV (Ly~$\alpha$) field at 10 AU in TW Hya, the closest young star with a circumstellar disk, is $\sim 10^{14}$ photons cm$^{-2}$ s$^{-1}$ \citep{herczeg_twhya1}, comparable to that used in the laboratory.    The UV radiation may also come from external high mass stars if a young solar-type star is born in a cluster.  Thus it is likely that -- on the disk surface -- that radicals are being created in the frozen mantle.
(2) Once the radicals have been generated then a cold ($\sim 10$ K) grain needs to be warmed to above the temperature of $\sim$25~K.     This could be possible if the grain absorbs an X-ray, cosmic ray,  undergoes a grain-grain collision, or advects to warmer layers.   

\subsubsection{Grain Surface Chemistry}
\label{sec:surchem}

A full treatment of grain surface chemistry is beyond the scope of this chapter and the reader is referred to the review by \citet{herbst05}.   There are a few issues specific to disk chemistry that should be discussed.   First at low dust temperatures ($< 20$ K) H atom addition will dominate as H (and D) atoms can more rapidly scan the surface when compared to heavier atoms or molecules.    The scanning timescale is \citep{tielens_book}:

\begin{equation}
\tau_m = \nu_m^{-1} e^{\left( -E_m/kT_{dust}\right)}.
\end{equation}

\noindent $\nu_m$ is the vibration frequency of the migrating molecules which is equivalent to $\nu_0$ (Eqn.~\ref{eq:nugr}).  It is typical to assume $E_m \sim 0.3 - 0.5E_b$, thus at $T_{dust} \sim 10$~K atoms with low binding energies (H, D) migrate rapidly and heaver molecules tend to remain in place.

The H atom abundance in molecular gas is set by the balance between H$_2$ ionization (which ultimately produces a few H atoms per ionization) and reformation of H$_2$ on grains.  Since the ionization rate in the disk midplane is likely small (see \S~\ref{sec:ion}) and cold grains are present, there may be few H atoms (or other atoms) present in the gas to allow for an active surface chemistry.  To be more explicit, the sequence of H formation and destruction reactions in a well shielded molecular medium are as follows:

\begin{eqnarray}
\rm{H  +   gr} &   \rightarrow & \rm{ H_{gr}}\\
\zeta_{\rm{CR,XR,RN}} \rm{+ H_2} & \rightarrow & \rm{H_2^+} + e\\
\rm{H_2^+ + H_2} & \rightarrow & \rm{H_3^+ + H}
\end{eqnarray}

The limiting step in H atom formation is the ionization from cosmic rays (CR), X-rays (XR), or radionuclides (RN).   
In essence every ionization of H$_2$  ultimately produces $f_H \sim$ 2--3 H atoms per ionization.  In equilibrium the H atom concentration is:

\begin{equation}
n_H = \frac{f_H\zeta n_{H_2}}{n_g \sigma_g v_H S_H\eta} = \frac{f_H \zeta}{10^{-21} v_H S_H \eta}.
\label{eq:hgr}
\end{equation}

\noindent In this equation $v_H$ is the thermal velocity of hydrogen atoms, and $\eta$ is an efficiency factor which can be set to unity provided that there is more than one H atom on each grain.
$S_H$ is the sticking coefficient for H atoms.   \citet{burke_hollenbach} and \citet{bz91} provide analytical expressions that can be used with values of $S_H \sim 0.7$ at 20 K.

We  have also used the expression for $n_g \sigma_g = 10^{-21}n$ discussed in \S~\ref{sec:gas-grain}.    Each of the relations in Eqn.~\ref{eq:hgr} are constant (given a temperature), thus the space density of H atoms is a constant in the molecular gas.  If we assume that cosmic rays penetrate to the midplane with an ionization rate of $\zeta = 1 \times 10^{-17}$ s$^{-1}$ (the most extreme high ionization case for the midplane) and the gas temperature is 30~K then $n_H \sim 5$ cm$^{-3}$.  The H atom abundance is therefore inversely proportional to the density.  The average grain in the ISM with a size of 0.1 $\mu$m has an abundance of $\sim 10^{-12}$.
This sets the initial conditions for the abundance of grains.  If $\eta = 1$ then for much of the disk midplane, where $n > 10^{12}$ cm$^{-3}$ there will be less than 1 H atom per grain significantly reducing H atom surface chemistry.  However, it may be the case that there is less than one H atom on each grain and $\eta \ll 1$ which could significant reduce the grain surface formation rate.  This would lead to more gas phase H atoms and a potential to power chemistry but requires longer timescales.

  Second, at higher temperatures heavier atoms and molecular radicals can begin to diffuse on the grain surface and create complex species.  Given the abundance of radiation on the disk surface it is highly likely  that bonds are broken in the ices creating radicals (from CH$_3$OH, H$_2$O, CO$_2$, CH$_4$, NH$_3$, ...).   Upon warm up this can produce more complex species \citep[see models of ][as an example]{gwh08}, unless it is limited by explosive desorption.  Thus grain chemistry may be important on the disk surface and, depending on the strength of vertical mixing, its products may reach the disk    midplane. 

\subsubsection{Deuterium Fractionation}
\label{sec:dfrac}

Enhancements of deuterium-bearing molecules relative to the hydrogen counterparts have been known for some time in solar system  and in the interstellar medium.   In cold ($T_{dust}$ $\sim$ 10--20~K) cores of clouds where stars are born, enrichments of 2-3 orders of magnitude are observed above the atomic hydrogen value of (D/H) $\geq (2.3 \pm 0.2) \times 10^{-5}$ estimated within 1 kpc of the Sun \citep{linsky06}.   Because of the lower zero point energy for deuterium compared to hydrogen, at low (T $< 30$ K) temperatures, it becomes favorable to transfer the D-bond, as opposed to the H bond.
 Ion-molecule reactions in the dense ISM are thought to be the mechanism responsible for these enrichments \citep{millar_dfrac}.    Deuterium fractionation can also be powered via reactions on the surfaces of dust grains \citep{tielens83,  nwk05}, which is also discussed below.
 
In the gas, deuterium chemistry is driven by the following reaction:

\begin{equation} \label{eq:h2dpform}
\rm{H}_3^+ + HD \leftrightarrow  \rm{H}_2D^+ + H_2 + 230 K.
\end{equation}
 
 \noindent  The forward reaction  is slightly exothermic favoring the production of \HtwoDp\ at 10 K, enriching the [D]/[H] ratio in the species that lie at the heart of  ion-molecule chemistry \citep{millar_dfrac}\footnote{Reaction~\ref{eq:h2dpform} has been measured in the lab at low temperatures finding that there is an additional dependence on the ortho/para ratio of \Htwo\ \citep{gdr02}.}.  These enrichments are then passed forward along reaction chains to  \DCOp , DCN,  HDO,HDCO, and others.

At densities typical of the dense interstellar medium ($n \sim 10^5$ \cc ), pure gas-phase models without freeze-out cannot produce significant quantities of doubly \citep[NHD$_2$;][]{roueff00} and triply deuterated ammonia \citep[ND$_3$;][]{lis02, vdt02}.
 This motivated a re-examination of the basic deuterium chemistry.  The primary advance in our understanding is two-fold: (1) deuteration reactions do not stop with \HtwoDp , rather they continue towards the formation of both \DtwoHp\ and \Dthreep , via a similar reaction sequence \citep{pv03, roberts_dfrac, walmsley_dep}:

\begin{equation}
 \rm{H}_2D^+ + HD \leftrightarrow  \rm{D}_2H^+ + H_2 + 180 K,
\end{equation}

\begin{equation}
  \rm{H}_2D^+ +  HD \leftrightarrow D_3^+ + H_2  + 230 K.
\end{equation}

\noindent (2) The freeze-out of heavy species in the disk midplane (\S~\ref{sec:gas-grain}), in particular CO, a primary destroyer of both \Hthreep\ and \HtwoDp , increases the rate of the  gas phase fractionation reactions. Thus in vertical gas layers, prior to potential complete freeze-out of heavy molecules, there can exist active deuterium fractionation, provided the gas temperature is below 30~K \citep[see][]{ah01}.

At high densities in the inner nebula thermodynamic equilibrium  can apply.  At temperatures  $T > 500$~K [HDO]/[H$_2$O] closely follows [HD]/[H$_2$], but at $T< 500$~K fractionation can proceed via the following reaction  \citep{rbj77}:

\begin{equation}
{\rm HD + H_2O \leftrightarrow HDO + H_2.}
\end{equation}

\noindent This reaction can provide a maximum enrichment of a factor of 3 \citep{lr92}.

Grain surface chemistry can also produce deuterium fractionation as the gas-phase chemistry (at low temperature) leads to an imbalance and enhancement in the D/H ratio of atomic H.  The efficiency of this process depends on the number of atoms which are generated by ionization.  
  At low temperatures more deuterium is placed in H$_2$D$^+$ relative to H$_3^+$.  These species create H and D atoms by reacting with neutrals.   Thus $f_D > f_H$ (see Eqn.~\ref{eq:hgr}) and there is the potential for these atoms to freeze onto grains and react, thereby fractionating molecular ices.
  However, as noted in \S~\ref{sec:surchem} it is possible that in the dense midplane that there will be less than one D atom generated per grain and the efficiency of D-fractionation on grain surfaces could be reduced.

 To summarize, deuterium chemistry can be active in the disk, both in the gas and on grains, but it requires low temperatures and sufficient levels of ionization.

\subsection{Kinetics and Thermodynamic Equilibrium}
\label{sec:valid}

At present it is not clear  exactly where in the disk thermodynamic equilibrium is valid.
It is clear that some aspects of meteoritic composition are consistent with equilibrium condensation (\S~\ref{sec:met}).   However, there are also aspects that are inconsistent, such as the detection of unaltered pre-solar grains \citep{hl95} and deuterium enrichments in cometary ices, which suggest that not all material cycled to high temperatures (above evaporation).
\citet{fegley_ssrv} provides a schematic of the relative importance of equilibrium to kinetics.  He proposes that thermodynamic equilibrium is of greater significance in the inner nebula and within giant planet subnebula.   To some extent this must be the case, but divining the exact transition is not entirely clear as, for example, the inner nebula and surfaces of giant planet subnebula will be exposed to radiation that powers non-equilibrium photochemistry.  Given evidence from trends in meteoritic composition  \citep[e.g.][as one example]{davis_geochem}, it is clear that in some instances thermochemical equilibrium is valid.  However, one must pay careful attention, where possible, to the relevant kinetic timescales.

\section{Physical Picture and Key Processes}

\subsection{Density and Temperature Structure}
\label{sec:pc}

Estimates of the physical structure of circumstellar disks can come from the solar system and also from extra-solar systems.   Within the solar nebula estimates are based on the so-called minimum mass solar nebula: the minimum amount of mass required to provide the current distribution of planetary mass as a function of distance from the Sun, correcting for a gas/dust ratio.   \citet{hayashi_mmsn} estimates $\Sigma (r) \sim 1700r^{-3.2}$, which is commonly used, but see also \citet{w77_mmsn} and \citet{davis_mmsn}.    It is important to note that this value is a lower limit and does not account for any potential solids that have been lost from the system during the early evolutionary phases.   Limits on the gas temperature have been estimated by \citet{lewis74}, based on the differences in planetary composition.  For instance the temperature at 1 AU is suggested to be $\sim 500-700$~K, based on the estimation that the Earth formed below the condensation temperature of FeS, but above the end point for oxidation of metallic iron in silicates \citep{lewis74}.   Based on these estimates the nebular thermal structure would decay as $R^{-1}$, which is much steeper than in pure radiative equilibrium $R^{-0.5}$ \citep[see, e.g.][]{hayashi_mmsn}.
However, as discussed below, radiation does dominate the thermal structure of protoplanetary disks.

 \begin{figure*}[t]
\centering
\includegraphics[width=12cm]{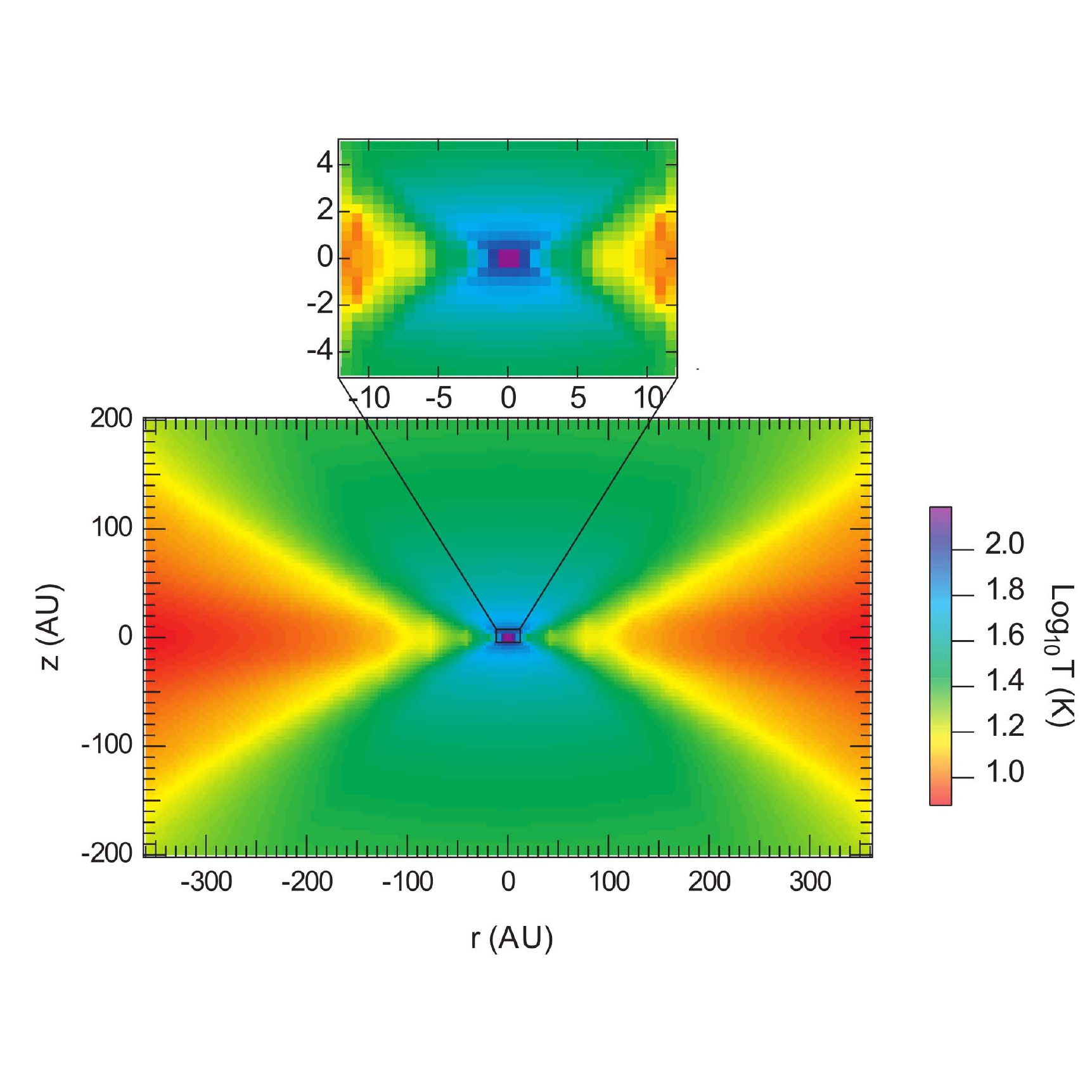}
 \caption{\small Predicted temperature structure for a typical T Tauri disk model by \citet{dch01}. 
 Assumed parameters are 0.5 M$_{\odot}$, $\dot{M} = 3 \times 10^{-8}$ $M_{\odot}$/yr, and $\alpha = 0.01$.
 \label{fig:tstruct}}
 \end{figure*}

More detailed estimates as a function of position (both radial and vertical) are available from extra-solar protoplanetary systems.  Models take into account the energy generated by accretion and by stellar irradiation of a flared disk structure. 
 Disk flaring is due to hydrostatic equilibrium \citep{kh87} and is characterized by a pressure scale height $H_p = \sqrt{R^3c_s^2/GM_*}$, with the sound speed, $c_s = \sqrt{kT/\mu m_H}$ ($\mu$ is the mean molecular weight,  $\mu \sim 2.3$).  The vertical density then follows $\rho = \rho_0 {\rm exp}(-z^2/2H_p^2)$, with $\rho_0$ a function of radius.    
  With radial  density profile as a variable and some assumptions regarding dust composition and size distribution (to set the optical properties) the spectral energy distribution of dust emission from UV/Visible ($\sim 4000$ \AA) to millimeter (1000 $\mu$m) can be predicted and compared to observations \citep{calvet91, cg97, dalessio98}.   These models have found that  disk surfaces, which are directly heated by stellar irradiation, are warmer than the midplane.   The midplane is heated by reprocessed stellar radiation  and, for the inner few AU,  by viscous dissipation of accretion energy.    
 Fig.~\ref{fig:tstruct} shows an example of the thermal structure from one such model \citep{dch01}.     
 
  An important issue  is the effect the initial steps of planetesimal formation, grain growth and settling, on the disk physical structure.  Smaller grains ($\langle a \rangle \sim 0.1$ $\mu$m) preferentially absorb the energetic short wavelength and as these grains grow to larger sizes their optical depth decreases \citep{dd04, dalessio06}.    The presence of small grains in the disk atmosphere leads to higher temperatures and greater flaring than disks with larger grains (or more settling to the midplane).   Thus the disk physical structure evolves with the evolution of solids.   Moreover there is an observed decay in accretion rate over similar timescales \citep{muz00}. Thus, the energy input from accretion should similarly decline.    In sum, both the midplane temperature and pressure should decline with evolution \citep{dcw05}.  As we will show in \S~\ref{sec:curr} this structure and evolution is an important component of time-dependent disk chemical models.

\subsection{Disk Ionization}
\label{sec:ion}

The gas-phase chemistry is powered by ionizing photons.  The characteristics of the chemical kinetics therefore depend on the flux and wavelength dependence of ionizing photons.   The molecular ions generated by photo- or cosmic ray absorption are also the key to linking the mostly (by many orders of magnitude) neutral gas to the magnetic field which is thought to have a direct relation to the physics of disk accretion.   In the following we will explore each of the various components of disk ionization.   In Table~\ref{tab:ion} we provide a listing of the major contributors to the ion fraction \citep[table taken from ][]{bergin_ppv}.   

\begin{sidewaystable}
\centering
\begin{threeparttable}[t]
\caption{Disk Ionization Processes and Vertical Structure\tnote{a}}
\begin{tabular*}{6.1in}[t]{lllll}
\hline
\hline
\multicolumn{1}{c}{Layer/Carrier} &
\multicolumn{1}{c}{Ionization Mechanism} &
\multicolumn{1}{c}{$\Sigma_{\tau = 1}$ (g cm$^{-2}$)\tnote{b}} &
\multicolumn{1}{c}{$\alpha_{r}$ (cm$^{-3}$s$^{-1}$)\tnote{c}} &
\multicolumn{1}{c}{$x_e$\tnote{d}} \\\hline
{\bf Upper Surface} & UV photoionization of H\tnote{e} &  6.9 $\times 10
^{-4}$
& $\alpha_{\rm{H}^+} = 2.5 \times 10^{-10}T^{-0.75}$ & $> 10^{-4}$\\
H$^+$ & $k_{\rm{H}^+} \sim 10^{-8}$ s$^{-1}$ &&\\\hline
{\bf Lower Surface} & UV photoionization of C\tnote{f} & 1.3 $\times 10^
{-3}$ &
$\alpha_{\rm{C}^+} =1.3 \times 10^{-10}T^{-0.61}$ & $\sim 10^{-4}$ \\
C$^{+}$ & $k_{\rm{C}^+} \sim 4 \times 10^{-8}$ s$^{-1}$ &&\\\hline
{\bf Warm Mol.} & Cosmic-\tnote{g} and X-Ray\tnote{h} Ionization
& 96 (CR) & $\alpha_{\rm{H_3}^+}=-1.3 \times 10^{-8} +$& $10^{-11\rightarrow -6}$  \\
H$_3^+$,HCO$^+$ & $\zeta_{cr} = \frac{\zeta_{cr,0}}{2}[{\rm exp}(-\frac{\Sigma_1}{\Sigma_{cr}}) +
{\rm exp}(-\frac{\Sigma_2}{\Sigma_{cr}})]$     & & 1.27 $\times 10^{-6}T^{-0.48}$ &
 \\
& $\zeta_X = \zeta_{X,0}
\frac{\sigma(kT_X)}{\sigma(1keV)}L_{29}J(r/{\rm AU})^{-2}$ & 0.008 (1keV) &  $\alpha_{\rm{HCO}^+} = 3.3 \times
10^{-5}T^{-1}$ &  \\
& & 1.6 (10 keV) &  &  \\\hline
{\bf Mid-Plane} & cosmic ray\tnote{g} and Radionuclide\tnote{i}&
 & & \\
& $\zeta_{R} = 6.1 \times 10^{-18}$ s$^{-1}$  & & &
 \\
Metal$^{+}$/gr & ($r < 3$ AU) & &
$\alpha_{\rm{Na}^+} = 1.4 \times 10^{-10}T^{-0.69}$ & $<
10^{-12}$ \\
HCO$^{+}$/gr & ($3 < r < 60$ AU) & & $\alpha_{gr}$ (see text) & 10$^{-13,-12}$ \\
H$_3^{+} -$ D$_3^+$ & ($r > 60$ AU) & & $\alpha_{\rm{D_3}^+} = 2.7 \times
10^{-8}T^{-0.5}$ & $> 10^{-11}$\\\hline
\end{tabular*}
 \begin{tablenotes}
\item [a] Table originally from \citet{bergin_ppv}
\item [b] Effective penetration depth of radiation (e.g., $\tau = 1$ surface).
\item [c]
Recombination rates from UMIST database \citep{rate99}, except for
H$_3^+$ which is from \citet{mccall06} and D$_3^+$ from \citet{larsson97}.
\item [d] 
Ion fractions estimated from \citet{swh04} and \citet{sano00}. Unless noted values are relevant for all radii.
\item [e]  Estimated at 100 AU assuming 10$^{41}$ s$^{-1}$
ionizing photons \citep{hyj_ppiv} and  $\sigma = 6.3 \times
10^{-18}$ cm$^{2}$ (H photoionization cross-section at threshold).
This is an overestimate as we assume all ionizing photons are at the
Lyman limit.
\item [f] Rate at the disk surface at 100 AU using the
radiation field from \citet{bergin_lyalp}.
\item [g] Taken from \citet{swh04}.  $\zeta_{cr,0} = 1.0 \times 10^{-17}$ s$^{-1}$ and
$\Sigma_1(r,z)$ is the surface density above the point with height $z$ and
radius
$r$ with $\Sigma_2(r,z)$ the surface density below the same point.  $\Sigma_{cr} =
96$ g
cm$^{-2}$ as given above \citep{un81}.
\item [h] X-ray ionization formalism from \citet{gfm_ppiv}.  $\zeta_{X,0} = 1.4 \times10^{-10}\;s^{-1}$, while
$L_{29} = L_X/10^{29}$ erg s$^{-1}$ is the X-ray luminosity and $J$ is an attenuation factor, $J =
A\tau^{-a}e^{-B\tau^b}$, where $A$ = 0.800, $a$ = 0.570, $B$ = 1.821, and $b$ = 0.287 (for energies around 1
keV and solar abundances).
\item [i] $^{26}$Al decay from \citet{un81}. If $^{26}$Al is not present $^{40}$K dominates with $\zeta_R = 6.9 \times 10^{-23}$ s$^{-1}$.
\end{tablenotes}
\end{threeparttable}
\label{tab:ion}
\end{sidewaystable}

\subsubsection{Cosmic Rays}

The interstellar cosmic ray ionization rate has been directly constrained by the Voyager and Pioneer spacecraft at $\sim 60$AU from the Sun \citep{webber_cosmicray} and estimates in the ISM have been summarized by \citet{dalgarno06}.   Based on these studies, the ionization rate is believe to be $\zeta_{H_2} \sim 5 \times 10^{17} s^{-1}$ in the dense ISM with growing evidence for higher rates in the diffuse gas \citep{mccall06, indriolo07}.  It is this rate that impinges on the surface of the disk.
\citet{un81} explored penetration depth of ionizing protons as a function of gas column and find that cosmic rays penetrate to a depth of $\Sigma \sim$ 96 g cm$^{-2}$.   Table~\ref{tab:ion} compares this e-folding depth to that of other potential contributors to the ion fraction.  Excluding radionuclides, cosmic rays
represent the best mechanism to ionize and power chemistry in the disk midplane.  However, if present, even cosmic rays will have difficulty penetrating to the midplane in the inner several AU's of disks around solar-type stars.  Another important question is whether ionizing cosmic rays are present in the inner tens of AU  at all.
Within our own planetary system, the Solar wind limits ionizing
cosmic rays to beyond the planet formation zone.  Estimates of mass loss rates from young star winds
significantly exceed the Solar mass loss rate \citep{dupree05}, and may similarly exclude high energy nuclei.     Cosmic rays will contribute to the chemistry for the outer disk and the detection of ions potentially provides some confirmation \citep{cd05}.

\subsubsection{Ultraviolet Radiation}
\label{sec:uv}

T Tauri stars are known to have excess UV flux much higher than their effective temperature,  T$_{eff} \sim$ 3000~K \citep{hg86}, which are  generated, at least in part, by accretion \citep{cg98}.
Observations have suggested that the UV radiation field (and X-ray ionization) play a key role in the observed molecular emission \citep{wl00, aikawa_vanz02}.   UV radiation is important as the primary molecular photodissociation bands lie below 2000 \AA\ and this radiation both dissociates molecules and can power a rich chemistry on the disk surface \citep[see, for example][]{th85}.

\begin{figure*}[t]
\centering
\includegraphics[width=11.5cm]{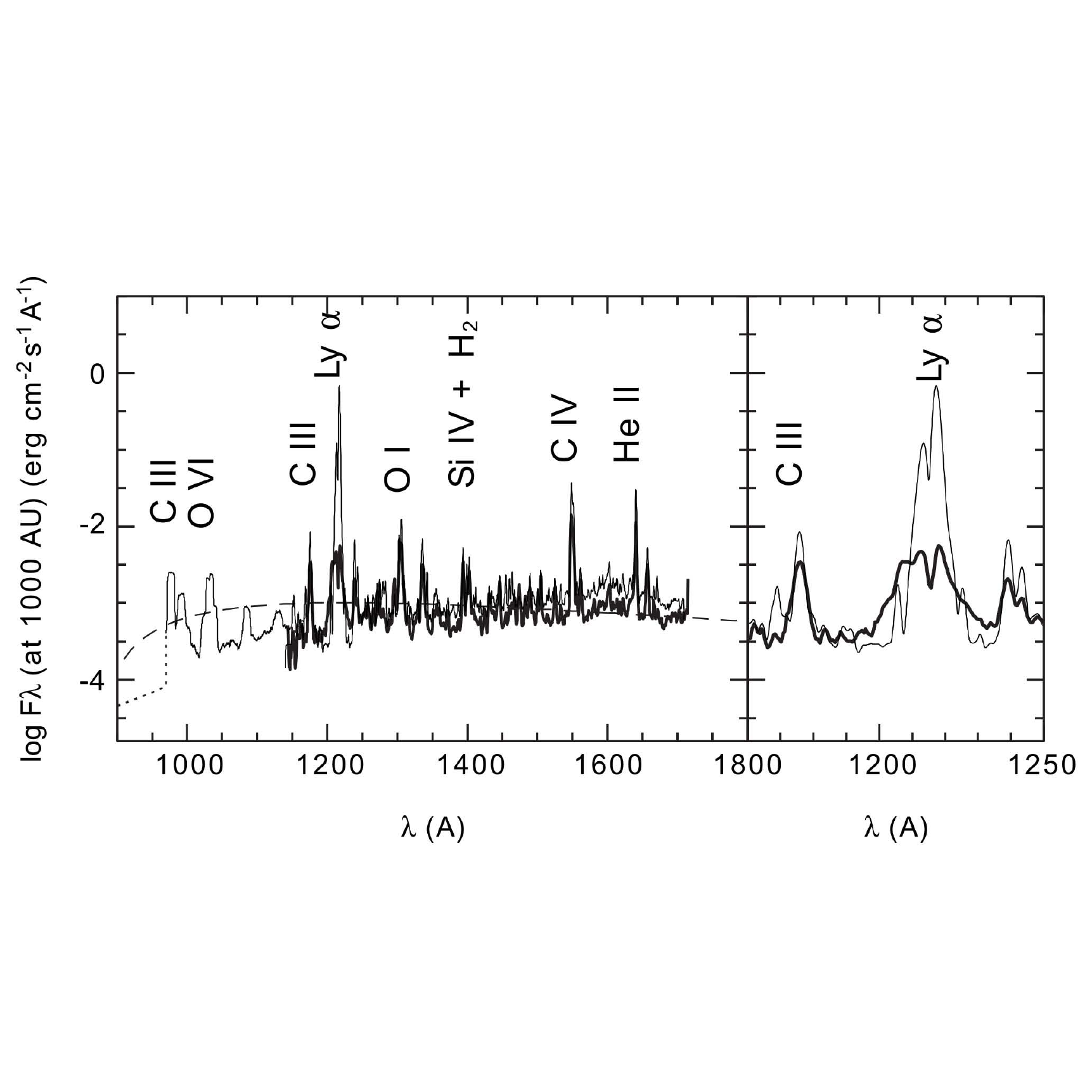}
 \caption{\small UV spectra of T Tauri stars. Heavy solid lines and light
solid lines represent the spectra of BP Tau and TW Hya, respectively.
The spectrum of TW Hya is scaled by 3.5 to match the BP Tau continuum level.
The long dashed line represents the interstellar radiation field of {\it Draine} (1978) scaled by a factor of 540.
The region around the Ly$\alpha$ line is enlarged in the right panel.
Taken from {\it Bergin et al.} (2003).
 \label{fig:uv}}
 \end{figure*}

In this context there are 2 photon fields to consider.  The stellar radiation field and the external radiation field which can be enhanced when compared to the standard interstellar radiation field (ISRF).   It is common to place both the stellar and interstellar UV radiation field in the context of the ISRF which is defined by the measurements of \citet{habing68} and \citet{draine78}.  
Using the measurements of Habing, the energy density of the ISRF is $1.6 \times 10^{-3}$ erg cm$^{-2}$ s$^{-1}$ which is defined as $G_0 = 1$ (in this context the Draine field is $G_0 = 1.7$).  As an example the FUV flux below 2000 \AA\ for T~Tau is $\sim 3 \times 10^{-13}$ erg cm$^{-2}$ s$^{-1}$ \AA$^{-1}$.
Scaled to 100 AU at 140 pc and integrated over 1000 \AA\ (the Lyman limit to 2000 \AA\ where most molecular photoabsorption cross-sections lie) gives 23.5 erg cm$^{-2}$ s$^{-1}$ or G$_0 = 1.5 \times 10^4$ at 100 AU (scaling as 1/r$^2$).    T~Tau has a particularly high mass accretion rate of dM/dt $\sim 10^{-7}$ M$_{\odot}$/yr that is about an order of magnitude higher than typically observed.  Most T Tauri stars have lower accretion rates and weaker UV fields with values in the range of $G_0(100\;\rm{AU}) = 300 - 1000$, based on observations from FUSE and HST \citep{bergin_h2}.  \citet{bergin_lyalp} pointed out the importance of Ly $\alpha$ emission, which is present in the stellar radiation field, but absent in the ISRF.  In one star that is unobscured by interstellar absorption, TW~Hya, Ly $\alpha$ contains as much of as 85\% of the stellar UV flux \citep{herczeg_twhya1}.  Even in systems where Ly $\alpha$ is absent due to absorption and scattering by interstellar \ion{H}{i}, Ly $\alpha$ pumped lines of H$_2$ are detected, testifying to the presence of  Ly $\alpha$ photons in the molecular gas \citep{herczeg_twhya1, bergin_h2, herczeg06}.  A sample of the UV field of typical T Tauri stars is shown in Fig.~\ref{fig:uv}.

\begin{figure}[t]
\centering
\includegraphics[width=10.5cm]{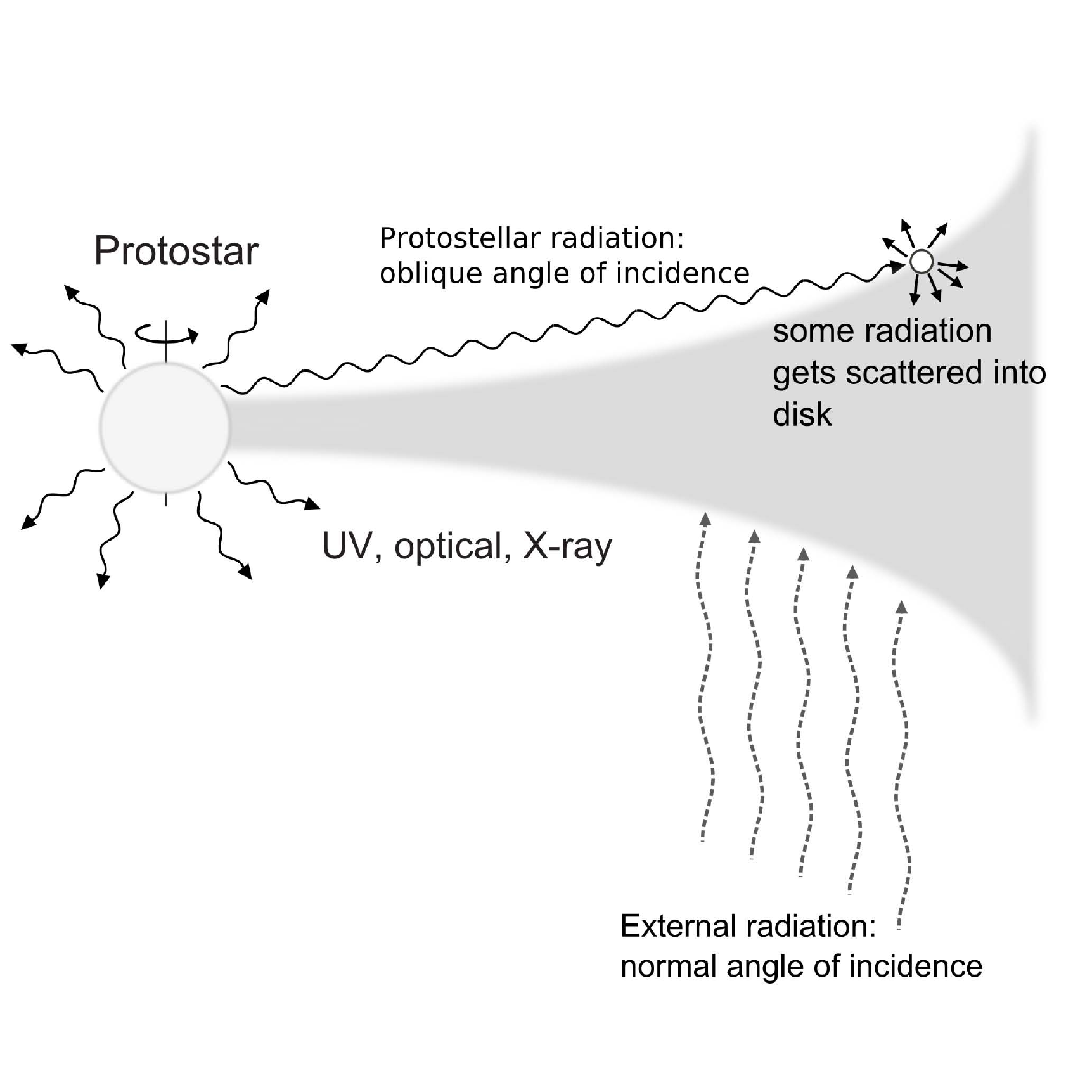}
\caption{Schematic showing the difference in angle of incidence between stellar and
external radiation impinging on a flared disk.   Radiation with a shallow angle of incidence  will have less penetration into the disk.   Scattering of stellar radiation increases its importance, particularly if the external radiation field suffers from any local extinction from a surrounding molecular cloud.
\label{fig:angle}}
\end{figure}

Stellar photons irradiate the flared disk with a shallow angle of incidence, while external photons impinge on the disk at a more normal angle.  Interstellar photons therefore have greater penetrating power \citep{wl00}.  \citet{vanz03} employed a 2D model of UV continuum photon transfer and demonstrated that scattering of stellar photons is key to understanding how UV radiation influences disk chemistry.  Fig.~\ref{fig:angle} illustrates the geometry of various contributors to the radiation field on the disk surface. 
An analytical approximation of the transfer of the stellar field is provided by \citet{bergin_lyalp}. In Taurus the average extinction towards T Tauri stars is A$_V \sim 1^m$ \citep{bergin_lyalp}; under these circumstances the stellar field will dominate over the ISRF ($G_0 = 1$) for much of the disk.  

Using the parameterization of the stellar field with respect to the ISRF allows the use of photodissociation rates calculated using the strength and shape of the IRSF, assuming albedo and scattering properties estimated for interstellar grains  \citep{vd88, roberge91}.   This correctly treats the direct attenuation of the field, but not the contribution from scattering, which dominates the deep surface \citep{vanz03}.  The shape of the stellar field, which is primarily line emission, is also different from the ISRF (see Fig.~\ref{fig:uv}).   This is particularly acute for Ly $\alpha$  as some species are sensitive to dissociation by Ly $\alpha$ emission (e.g. H$_2$O, HCN), while others are not, such as CO and CN  \citep[][]{bergin_lyalp, vd06}.
In this case, it is better to directly calculate the photodissociation and ionization rates from the flux and the cross-sections.   This can be done using the observed UV field, or a proxy black-body, in the following fashion:

\begin{equation}
k_{photo} = \int \frac{4\pi \lambda}{hc} \sigma(\lambda ) J_{\lambda,0}(r) e^{-\tau_{\lambda}} d\lambda\;(\rm{s}^{-1}).
\end{equation}

\noindent $\sigma(\lambda )$ is the photodissociation cross-section and $J_{\lambda, 0}$ is the mean intensity of the radiation field at the surface ($F_{\lambda} = 4\pi J_{\lambda} = \int I_{\lambda} d\Omega$, where $\Omega$ is the solid angle of the source).    The main source of opacity at these wavelengths is dust grains and some assumptions must be made regarding the composition, optical properties, and size distribution (see \S 4.3 for a more complete discussion of dust grain absorption of radiation).   \citet{vd06} provides photo-rates for various blackbodies and also photodissociation cross-sections coincident with Ly $\alpha$.  In addition, a web site for cross-sections is available at \url{http://www.strw.leidenuniv.nl/~ewine/photo}.  

 It is important to note that the photodissociation of two key species, H$_2$ and CO, require additional treatment.  These species are dissociated via a line-mediated process, as opposed to a dissociation continuum  typical for most other molecules.   In this regard the optical depth of the lines depends on the total column.   When the lines become opaque molecules closer to the radiation source can shield molecules downstream in a process that is called self-shielding.  In addition, H$_2$ molecules can partially shield some of the CO predissociation lines.  Self-shielding rates are time and depth dependent.  In general, approximations based on the interstellar UV field are adopted from \citet{db96}, for H$_2$, while for CO the shielding functions provided by \citet{vdb88} and \citet{lee_coh2} can be used.  One caveat of these approximations is that the shielding functions are calculated assuming that molecules predominantly reside in the ground state with a line profile appropriate for cold interstellar conditions.   In regions with high temperatures these approximations can potentially break down.

\subsubsection{X-ray Ionization}
 
 X-rays are a key source of ionization and heating of protoplanetary disks.    The typical X-ray luminosity (based on protostars in Orion) is $L_x = 10^{28.5} - 10^{31}$ ergs s$^{-1}$ with a characteristic temperature of the X-ray emitting plasma of $T_x \sim 1 - 2$ keV \citep[e.g.][see also chapter by Feigelson, this volume]{fm99}.   X-ray photons interact directly with the atoms in the disk; i.e., the inclusion of the atom in a molecule does not affect the photoabsorption cross-section.  In particular, the opacity is dominated by inner shell ionization of heavier elements.    For atoms heaver than Li the inner shell ionization is followed by the Auger effect which generates a number of primary electrons.   Each primary electron ($p$) produces secondary electrons ($s$) by impact ionization of the gas, with $N_{s} \gg N_{p}$.  Therefore for the X-ray ionization rate, $\zeta_x$:
 
 \begin{equation}
 \zeta_x = \zeta_{p} + \zeta_{s} = \zeta_{s}.
 \end{equation}
 
 \noindent This is a well known result and is similar to that seen for cosmic rays.   In general there are $N_s \sim$ 30 atoms and molecules ionized per keV \citep{ah01}.  
  
\begin{table}[t]
\centering
\begin{threeparttable}
\caption{Coefficients for Fit to X-ray Cross-Section\tnote{a,b}}
\begin{tabular}{lrrr}
\hline\hline
\multicolumn{1}{c}{Energy Range (keV)} &
\multicolumn{1}{c}{$c_0$} &
\multicolumn{1}{c}{$c_1$} &
\multicolumn{1}{c}{$c_2$}\\\hline\hline
0.030$-$0.100 & 17.3 & 608.1 & -2150.0 \\
0.100$-$0.284 & 34.6 & 267.9 & -476.1 \\
0.284$-$0.400 & 78.1 & 18.8 & 4.3\\
0.400$-$0.532 & 71.4 & 66.8 & -51.4\\
0.532$-$0.707 & 95.5 & 145.8 & -61.1 \\
0.707$-$0.867 & 308.9 & -380.6 & 294.0\\
0.867$-$1.303 & 120.6 & 169.3 & -47.7\\
1.303$-$1.840 & 141.3 & 146.8 & -31.5\\
1.840$-$2.471 & 202.7 & 104.7 & -17.0\\
2.471$-$3.210 & 342.7 & 18.7 & 0.0\\
3.210$-$4.038 & 352.2 & 18.7 & 0.0\\
4.038$-$7.111 & 433.9 & -2.4 & 0.75\\
7.111$-$8.331 & 629.0 & 30.9 & 0.0\\
8.331$-$10.000 & 701.2 & 25.2 & 0.0\\\hline
\label{tab:xray}
\end{tabular}
 \begin{tablenotes}
 \item [a] Fit and tabulation reprinted from \citet{mm83}.
 \item [b] Cross-section per hydrogen atom, $\sigma_T (E)$,  is ($c_0 + c_1E + c_2 E^2)E^{-3} \times 10^{-24}$ cm$^2$ (E in keV).
\end{tablenotes}
\end{threeparttable}
\end{table}

 The X-ray ionization rate can be derived by an integration of the X-ray flux, $F(E)$, and the cross-section $\sigma_i(E)$ of element $i$ (in this case hydrogen)  over energy:
 
 \begin{equation}
 \zeta_{x,i} = N_s \int_{1\;{\rm keV}}^{30\;{\rm keV}} \sigma_i (E) F(E)dE
 \end{equation}
 
\noindent The X-ray flux can be estimated from the X-ray luminosity assuming a characteristic temperature for the plasma \citep[see][for an example]{glassgold97}.  The X-ray flux decays with depth into the disk, $F(E) = F_{X,0} e^{-\tau (E)}$, where $\tau (E) = N\sigma_T(E)$.  
X-ray photons are attenuated by photoabsorption at the atomic scale.   In this regard is customary to use the total X-ray cross-section, $\sigma_T (E)$,  computed using solar abundances \citep{mm83}.  With this  assumption  the total hydrogen column is the only information required.  Table~\ref{tab:xray} list the fit to the X-ray cross-section computed by \citet{mm83}.
This assumes that the heavy elements, that control the opacity, have Solar abundances and are uniformly distributed in the disk.  

 It is well known that dust grains suspended in the disk atmosphere will sink to the midplane, reducing the dust/gas mass ratio in the upper disk atmosphere \citep{wc_ppiii}.    This will reduce the optical depth of the upper layers to both X-rays and UV radiation.  Compared to UV transfer, dust coagulation will not  reduce the X-ray opacity as dramatically, unless grains grow to large sizes.  Instead the mass of heavy elements must be redistributed as is the case of dust settling.   In the extreme limit only H and He are present in the upper atmosphere and the cross-section at 1 keV is reduced by a factor of $\sim$4.5 \citep[i.e. the H and He absorption limit,][]{mm83}.   In the case of non-solar abundances or mass redistribution the cross-section for each individual element can be used \citep{henke93}.
  X-ray ionization is crucial for the inner disk where cosmic rays likely do not penetrate.  X-ray photons also have a significantly greater penetration depth than UV photons and thus can power chemistry \citep[and perhaps accretion,][]{gammie96} on the disk surface.  The characteristics of the X-ray driven chemistry can extend beyond the simple ionization of H$_2$ as X-rays can produce short-lived, but reactive, doubly ionized molecules.  For greater detail, the reader is referred to \citet{langer78}, \citet{kk83},  and \citet{stauber05}. 

\subsubsection{Active Radionuclides}

In the case that the midplane of the disk is completely shielded from cosmic ray radiation then radionuclides provide a baseline level of ionization.   A primary issue in this regard is the overall abundance of various radionuclides and whether these elements are uniformly distributed throughout the system.   Key elements in this regard are $^{26}$Al and $^{60}$Fe which have short lifetimes [$\tau_{1/2}(^{26}{\rm Al}) = 0.72$ Myr; $\tau_{1/2}(^{60}{\rm Fe}) = 1.5$ Myr] and have been inferred to be present in the solar system \citep[e.g.][]{wadhwa_ppv, wasserburg06}.  If present they are a strong source of ionization for several millions of years.   Table~\ref{tab:ion} provides an estimate of this ionization rate, which is dominated by $^{26}$Al.  This is sufficient to allow for an active ion driven gas chemistry in regions where neutral molecules (excluding H$_2$ and HD) are not completely frozen on grains.  This ionization rate will decay with time according to respective half-life.    If $^{60}$Fe or $^{26}$Al are not present then this rate gets substantially reduced.
 \citet{fg97} provide a short summary of the relevant ionization processes including the most important radionuclides.

\subsubsection{Dissociative Recombination}

A key to the question of the overall ionization fraction is the balance between the ionization rate and the recombination rate of atoms and molecules.   In equilibrium, the ion fraction, $x_e = n_e/n$, can be expressed by:
$x_e = \sqrt{\zeta/(\alpha_{r} n})$,
where $\alpha_{r}$ is the
electron recombination rate.  The dependence of the ion fraction on position, therefore not only depends
on the flux of ionizing agents, but also on the recombination rate of the most abundant ions.   There are some key differences in this regard that are summarized in Table~\ref{tab:ion}.   For the midplane a key question is whether molecular ions, metal ions, or grains are the dominant charge carriers and this serves as a useful foil to discuss some differences in recombination.

In general atomic species have longer recombination timescales than molecular ions.   Thus the presence of abundant slowly recombining metals can have key consequences for magnetic field coupling \citep{in06b}.  However, metal ions are not present in the dense star forming cores that set the initial conditions for disk formation \citep{mb07}, and are likely not present in the disk gas.
If grains are the dominant charge carrier the recombination rate, $\alpha_{gr}$, is the grain
collisional timescale with a correction for long-distance Coulomb focusing:
$\alpha_{gr} = \pi a_d^2 n_{gr}v(1 + e^2/ka_dT_{dust})$. 
At $T_{dust} = 20$ K, \citet{ds87}
show that for molecular ions, grain recombination will dominate when
$n_e/n < 10^{-7}(a_{min}/3$\rm{\AA}$)^{-3/2}$.
Grains can be positive or negative and carry multiple charge: \citet{sano00} find that the
total grain charge is typically negative, while the amount of charge is 1-2e$^-$, varying with
radial and vertical distance.  \citet{fm06} provides a summary of molecular ion recombination rates determined in the laboratory.

\subsection{Grain Growth and Settling}

The onset of grain evolution within a protoplanetary disk
consists of collisional growth of sub-micron sized
particles into larger grains; the process continues until the larger grains
decouple from the gas and settle to an increasingly dust-rich midplane
\citep{nnh81, wc_ppiii}.

Grain coagulation can alter the chemistry through the reduction in the total
geometrical cross-section, lowering the adsorption rate and the Coulomb
force for ion-electron grain recombination.
Micron-sized grains couple to the
smallest scales of turbulence \citep{wc_ppiii}
 and have a thermal,
Brownian, velocity distribution. Thus, the timescale of grain-grain collisions is,
$\tau_{gr-gr} \propto a_d^{5/2}/(T_{dust}^{1/2}\xi n)$, $\xi$
the gas-to-dust mass ratio, and $T_{dust}$ the dust temperature \citep{aikawa99}.
In this fashion grain coagulation proceeds faster at small radii where the
temperatures and densities are higher.
\citet{aikawa99} note that the longer
timescale for adsorption on larger grains leaves more time for
gas-phase reactions to drive toward a
steady-state solution; this involves more carbon trapped in CO 
as opposed to other more complex species.

Overall, the evolution of grains, both coagulation and sedimentation,
can be a controlling factor for the chemistry.  As grains grow and settle to the midplane, the
UV opacity, which is dominated by small grains,
decreases, allowing greater penetration of ionizing/dissociating photons.
As an example, in the coagulation models of \citet{dd04}  the
integrated vertical UV optical depth at 1 AU decreases over several orders of
magnitude, towards optically thin over the entire column \citep[see also][]{weidenschilling97}.

This can be understood by exploring the question of dust extinction of starlight from the perspective of the interstellar medium.   In this case the amount of dust absorption is treated in terms 
of magnitudes of extinction, labelled as $A_{\lambda}$.\footnote{Magnitudes in astronomy are defined such that a difference of 5 mag between 2 objects corresponds to a a factor of 100 in the ratio of the fluxes.   Thus an extinction of 1 mag corresponds to a flux decrease of $(100)^{1/5}  \sim 2.512$.}  
At any given point in the disk (or in the ISM) the intensity can be charactered by the intensity (ergs s$^{-1}$ cm$^{-2}$ \AA$^{-1}$ sr$^{-1}$) that impinges on the surface,  $I_{\nu, 0}$, modified by the dust absorption, $I_{\nu} = I_{\nu,0}{\rm exp}(-\tau_{\lambda})$.   The opacity,  $\tau_{\lambda}$, is given by 
$\tau_{\lambda} = N_{dust} Q_{\lambda} \sigma $, where $N_{dust}$ is total dust column along the line of sight, $Q_{\lambda}$ is the extinction efficiency, and $\sigma$ is the geometrical cross-section of a single grain.
The wavelength dependent extinction, placed in terms of magnitudes, is  defined below.

\begin{eqnarray}
A_{\lambda} & =& -2.5{\rm log}(I_{\nu}/I_{\nu, 0}) \\
 &=& 2.5 {\rm log}(e) \tau_{\lambda} \\
  &=& 1.086 N_{dust} Q_{\lambda} \sigma \\
  &=& 1.086 \frac{n_{dust}}{n} N Q_{\lambda} \sigma \label{eq:al}
\end{eqnarray}

\noindent The total dust column can be related to the gas column ($N$) via the gas to dust mass ratio, $\xi$, which is measured to be 1\% for interstellar grains (i.e. the primordial condition):

\begin{equation}
\frac{\rho_{dust}}{\rho_{gas}} = \xi = \frac{n_{d}m_{gr}}{n \mu m_H},\\
\end{equation}

\begin{equation}
\frac{n_{d}}{n} = \frac{3}{4\pi} \frac{\xi \mu m_H}{\rho_{gr} a_{gr}^3}.
\end{equation}

\noindent  Placing the dust abundance ($n_{dust}/n$) into Eqn.~\ref{eq:al} and exploring the extinction at visible wavelengths where $Q_V \sim 1$, assuming a typical grain size of 0.1 $\mu$m, and $\rho_{gr} = 2$ g cm$^{-3}$ then,

\begin{equation}
A_V = 1.086  \frac{3}{4} \frac{\xi \mu m_H}{\rho_{gr} a_{gr}} N Q_{V} \sim \left(\frac{\xi}{0.01}\right)\left(\frac{Q_V}{1}\right) 10^{-21}N.
\end{equation}

 \noindent Thus a total gas column of $10^{21}$ cm$^{-2}$ provides 1 mag of extinction at visible wavelengths, for a standard gas-to-dust mass ratio. As grains settle to the midplane the gas remains suspended in the upper layers and the gas-to-dust ratio decreases.   This requires larger columns for the dust to absorb stellar radiation.
 In our example we have explored extinction in the visual, while most molecules are dissociated by ultraviolet radiation.   Dust extinction is larger in the UV (requiring a smaller gas column); however, the dependency on the gas-to-dust ratio and grain size remains.
 
 As grains evolve there will be a gradual
shifting of the warm molecular layer deeper into the disk, eventually into the
midplane \citep{jonkheid_tgas}.  Because the grain emissivity, density, and temperature will also change, the
chemical and emission characteristics of this layer may be altered 
\citep{an06}.
These effects are magnified in the inner disk, where there is
evidence for significant grain evolution in a few systems \citep{furlan_graingrowth}
and deeper penetration of energetic radiation \citep{bergin_h2}.
A key question in this regard is the number of small grains (e.g., PAHs)
present in the atmosphere of the
disk during times when significant coagulation and settling has occurred.

\subsection{Turbulence and Mixing}

It is becoming increasingly clear that mixing played an important role in the chemistry of at least the solids in the nebula.  This is due to the detection of crystalline silicates in comets \citep[e.g.][]{brownlee_stardust} and the presence of chondritic refractory inclusions in meteorites \citep{macp88}.  Since the smallest solids (micron size grains) are tied to the gas it is likely that the gas is also affected to some extent (perhaps in a dominant fashion) by turbulent mixing.   The question of mixing and its relevance has a long history in the discussions of solar nebula chemistry \citep[e.g.][and references therein]{prinn_ppiii}.

In terms of the dynamical movement of gas within a protoplanetary disk and its
chemical effects, a key question is whether the chemical timescale, $\tau_{chem}$, is less
than the relevant dynamical timescale, $\tau_{dyn}$, in which case  the chemistry will
be in equilibrium and unaffected by the motion.
If $\tau_{dyn} <  \tau_{chem}$ then mixing will alter
the anticipated composition.  These two constraints are the equilibrium and
disequilibrium regions (respectively) outlined in  \citet{prinn_ppiii}.
What is somewhat different in our current perspective is the recognition of an active
gas-phase chemistry on a photon-dominated surface (\S~\ref{sec:curr}).
This provides another potential mixing reservoir in the vertical direction, as
opposed to radial, which was the previous focus.


It is common to parameterize the transfer of angular momentum
in terms of the turbulent viscosity,
$\nu_t = \alpha c_s H_p$,
where $\nu_t$ is the viscosity, $c_s$ the sound speed, $H_p$ the disk scale
height, and $\alpha$ is a dimensionless parameter \citep{ss73, pringle81}.
\citet{hartmann98} empirically constrained the $\alpha$-parameter to be
$\lesssim
10^{-2}$ for a sample of T Tauri disks.

A number of authors have begun to explore the question of including
dynamics into the chemistry.   A brief and incomplete list is provided by \citet{bergin_ppv}. 
While the details differ, a common approach is to use mixing length theory where the transport is treated as a diffusive process \citep{ayw81, xal95}, but see also \citet{ilgner04, tg07}.   For mixing length models the fluctuations of abundance of species $i$ can be determined by the product of the abundance fluctuations in a given direction and the mixing length ($l$), $dx_i \sim -l dx_i/dz$, where $z$ denotes the direction where a gradient in abundance exists.   The net transport is then \citep{willacy06}:

\begin{equation}
\phi_i ({\rm cm}^{-2}\;{\rm s}^{-1}) = n_{{\rm H}_2} \langle v_t dx_i \rangle = -Dn_{{\rm H}_2} \frac{dx_i}{dz} = -Dn_i \left[\frac{1}{n_i}\frac{dn_i}{dz} - \frac{1}{n_{{\rm H}_2}}\frac{dn_{{\rm H}_2}}{dz}\right],
\end{equation}

\noindent where $D$  is the diffusion coefficient and $v_t$ is the turbulent velocity.  The diffusivity is related to the viscosity $\nu_t$ by $D =  \langle v_t l\rangle = \nu_t = \alpha c_s H_p$.  Using this description,  the chemical continuity equation can be written as:

\begin{equation}
\frac{\partial n_i}{\partial t} + \frac{\partial \phi_i}{dz} = P_i - L_i,
\end{equation}

\noindent where $P_i$ and $L_i$ are the chemical production and loss terms for species $i$.

The radial disk viscous timescale is
$\tau_{\nu} = r^2/\nu$ or,

\[
\tau_{\nu} \sim 10^{4}{\rm yr}\left(\frac{\alpha}{10^{-2}}\right)^{-1}
\left(\frac{T_{gas}}{100\,{\rm K}}\right)^{-1}\left(\frac{r}{1\,{\rm AU}}\right)^{\frac{1}{2}}
\left(\frac{M_*}{M_{\odot}}\right)^{\frac{1}{2}}.
\]

\noindent The diffusivity, $D$ or $K$, is not necessarily the same as the
viscosity, $\nu_t$ \citep[e.g.][]{stevenson90}, as given above.    Moreover, it is not entirely clear that radial mixing will be the same as vertical mixing.  In the case of disks, several groups have explored this 
question with a range of potential solutions ranging from $\nu_t/D$ below unity to near 20
\citep{csp05, jkm06, turner06, yd07}.
 For completeness, when $\nu_t/D > 1$  turbulent mixing is much less efficient
than angular momentum transport.

\section{Current Understanding}
\label{sec:curr}

In this section we will synthesize the various physical and chemical processes in a discussion of our evolving understanding of disk chemistry.  One important issue is that because of its low binding energy to grain surfaces \citep{hs71}, molecular hydrogen will remain in the gas throughout the nebula.    
Provided that sufficient levels of ionizing agents reach the midplane or the deep interior of the disk surface then the ion-molecule chemistry outlined earlier can proceed.   Surface chemistry based on diffusing hydrogen atoms will be more difficult to initiate, but in warmer regions heavier radicals can potentially migrate and react.  Thus both gas and grain surface chemistry is important.

\begin{figure}[t]q
\centering
\includegraphics[width=12cm]{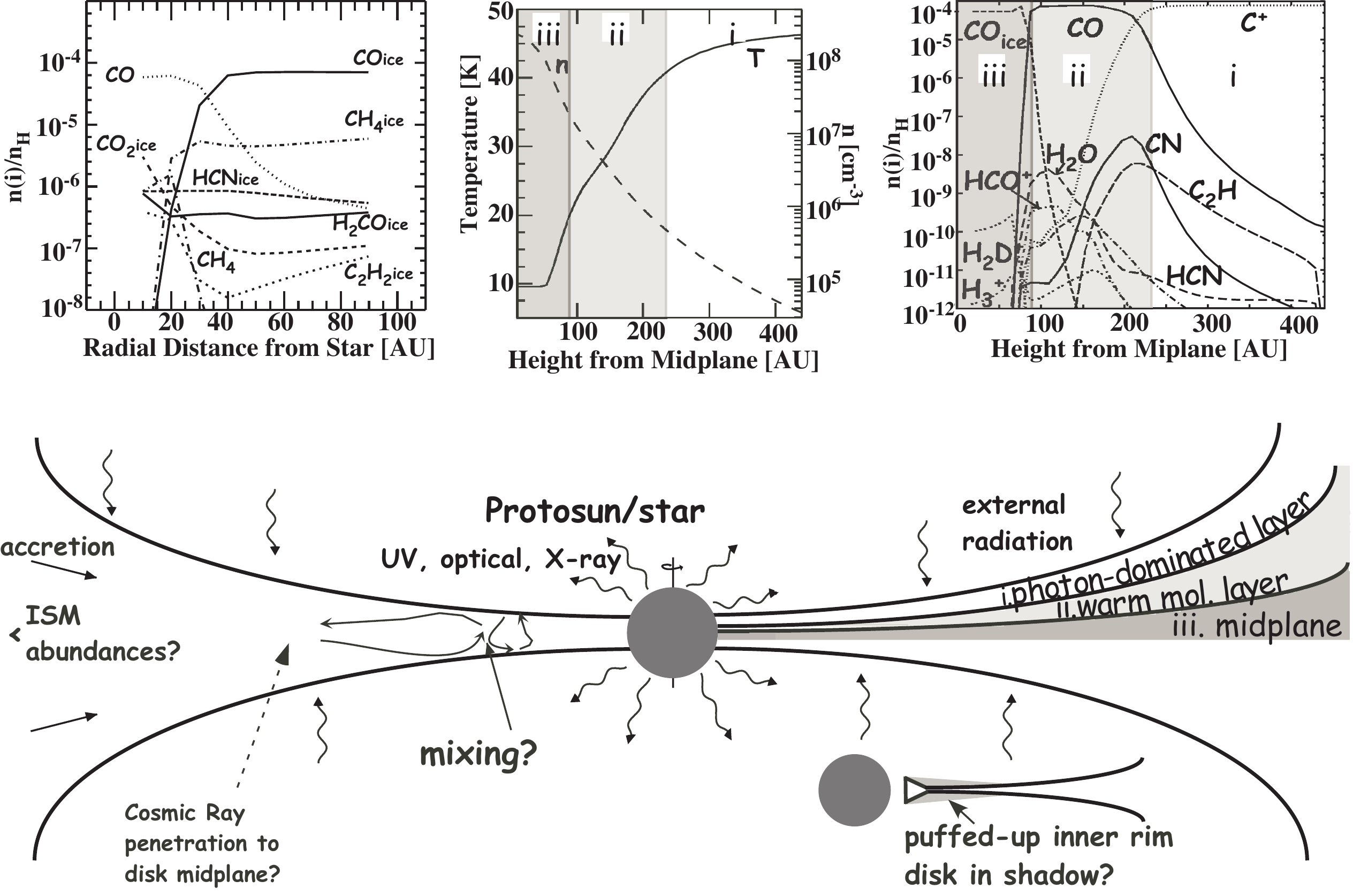}
 \caption{\small Chemical structure of protoplanetary disks \citep[taken from][]{bergin_ppv}. Vertically the disk is
schematically divided into three zones: a photon-dominated layer, a warm molecular
layer, and a midplane freeze-out layer. The CO freeze-out layer disappears at
$r\lesssim 30-60$ AU as the mid-plane temperature increases inwards. Various non-thermal inputs,
cosmic ray, UV, and X-ray drive chemical reactions. Viscous accretion and
turbulence will transport the disk material both vertically and radially.
The upper panels show the radial and vertical distribution of molecular
abundances from a typical disk model at the midplane \citep{aikawa99} and $r\sim 300$ AU \citep{vanz03}.  A sample of the hydrogen density and {\em dust} temperature at the
same distance \citep{dalessio99} is also provided.  
}
\label{fig:overview}
\end{figure}

For other molecules we expect large compositional gradients in the vertical and radial directions.  The dominant effect is the sublimation/freeze-out as molecules transition from warm to colder regions in a medium with high densities ($n \sim 10^{7} - 10^{15}$ cm$^{-3}$) and therefore short collisional timescales between gas and solid grains.   In large part the overall chemical structure follows the thermal structure and is schematically shown in Fig.~\ref{fig:overview}.    This plot illustrates that, beyond $\sim$40 AU, the disk can be divided into three vertical layers. 
CO, as one of the most volatile and abundant species, controls the gas-phase chemistry in the outer disk  and sets the radial and vertical boundary of these layers.    The top of the disk is dominated by stellar UV and X-ray radiation which leads to molecular photodissociation in a photon-dominated layer.   
This layer will have several transitions.   H/H$_2$ is the first transition, as molecular hydrogen is the strongest at self-shielding.  This is followed by \ion{C}{ii}/\ion{C}{i}/CO and subsequently the oxygen/nitrogen pools which require dust shielding \citep[e.g. \ion{O}{i}/H$_2$O, \ion{N}{i}/N$_2$ or NH$_3$; see][]{hollenbach_rvmp}.

 Inside of this photon-dominated layer, the grain temperatures are warm enough for CO to exist in the gaseous state ($T_{dust} > 20$~K), but water will remain as an ice.   Thus, the C/O ratio $\sim$ 1 leading to an active carbon-based chemistry in a vertical zone labeled as the ``warm molecular layer''.   The ion-molecule chemistry of this layer is powered predominantly by X-rays and Cosmic Rays (if present).   Reactions between CO and molecular ions (predominantly H$_3^+$) transfer a small fraction of this carbon into other simple and complex species  \citep{aikawa97}.   If CH$_4$ is present on grain surfaces, and evaporates into the gas, it will also be a key pre-cursor to the creation of larger hydrocarbons and carbon chains.   If  $T_{dust} < T_{sub}$ for  any product of this gas-phase chemistry, then that molecule freezes onto grains.  In a large sense the gas-phase is acting as an engine for building complexity within molecular ices \citep{tg07}.  As material advects inwards the region of the disk with $T_{dust} > 20$~K increases in depth and, in addition, the more tightly bound complex species created earlier (e.g. HCN, C$_2$H$_2$, C$_3$H$_4$)  will eventually evaporate.
   An additional issue that is likely important is the creation of frozen radicals via photodissociation on the grain surface.   If the grain is warm enough these radicals can migrate and react  on grain surfaces.  This can lead to a large increase in molecular complexity \citep{gwh08}, perhaps contributing to organics detected in meteorites \citep[see, e.g.][and references therein]{2002ApJ...576.1115B}.
   
   Below the warm molecular layer lies the dense ($n \gg 10^8$ cm$^{-3}$), cold ($T \le 20$~K) midplane where molecules are frozen on grain surfaces \citep{wl00, ah01}.  This is the case for much of the outer disk (R $\gtrsim$ 20 - 40~AU) beyond the radial ``snow-lines''.   If ionizing agents are still present and $T_{gas} < 30$~K,  then the transition to the midplane and the  midplane itself are the main layers for deuterium fractionation in the disk \citep{ah01}.   Gas phase deuterium fractionation requires heavy molecules to be present in the gas.  In the limit of total heavy element (C, O, N) freeze-out, the chemistry will be reduced to  the following sequence H$_3^+$ $\rightarrow$ H$_2$D$^+$ $\rightarrow$ D$_2$H$^+$ $\rightarrow$ D$_3^+$ \citep{cd05}.

\begin{figure}
\centering
\includegraphics[width=8cm]{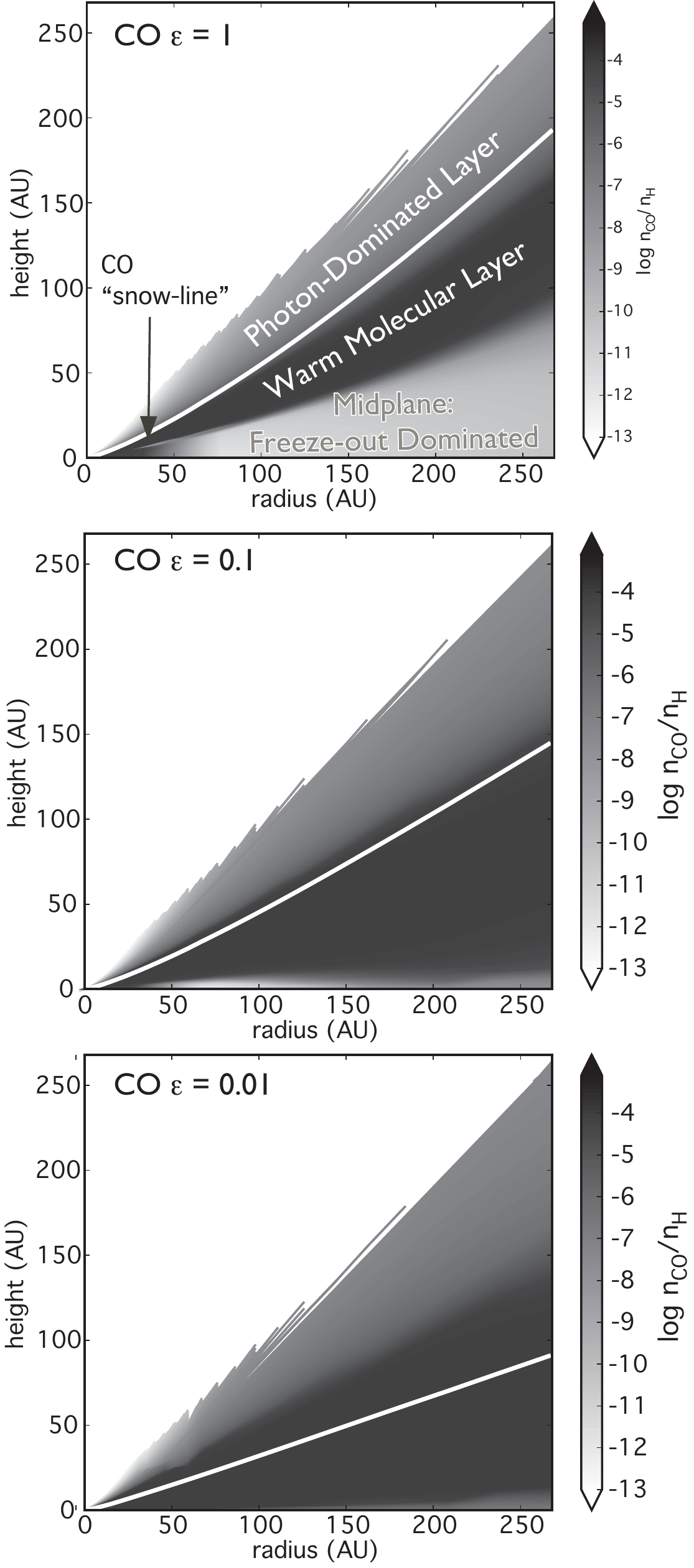}
 \caption{\small  Plot of the abundance of carbon monoxide as a function of radial and vertical height in three disk models with variable gas to dust ratios.   The gas to dust ratio is parameterized by the parameter $\epsilon$, which is the gas to dust ratio in the upper layers relative to the gas to dust ratio in the midplane (see Calvet, this volume).    Lower values of $\epsilon$ are representative of the effects of the settling of dust grains.   Also shown is the $\tau = 1$ surface for stellar photons which appears deeper in the disk as grains settle to the midplane.   The top panel shows some of the basic disk chemical structure.
 Figure from Fogel et al. 2009, in preparation. 
}
\label{fig:co}
\end{figure}
 
Fig.~\ref{fig:co} shows the CO abundance structure in a fiducial disk model to
 illustrate some of these effects more directly.  The three panels refer to different models of
 dust settling, with lower values of the $\epsilon$ parameter (defined in the figure caption) referring
 to a greater degree of dust settling.    The top panel is for a disk with no settling.  As can be seen the chemical structure
 divides into three layers as shown in Fig.~\ref{fig:overview}.  In addition the interior evaporation zone (inside the snow-line for
 CO) can be resolved.   As grains settle into the disk the radiation penetrates more deeply as seen by the line delineating the $\tau = 1$ surface and also in the chemical structure which moves the warm layer to lower depths \citep{jonkheid_tgas}.

 The case of water deserves special mention.   Beyond the snow-line at $\sim 1 - 5$ AU, water will exist mostly in the form of water ice \citep[see][]{cc06,davis07}.   Dust grains suspended in the atmosphere will typically not be heated above the evaporation temperature (Fig.~\ref{fig:tsub}) and water will, for the most part, remain as ice.
However, UV photons can photodesorb water ice from grain surfaces, producing a small layer where water exists with moderate abundance \citep[$x_{H_2O, max} \sim 10^{-7}$, relative to H$_2$;][]{dchk05}.  The water vapor abundance peaks at the layer where the rate of photodesorption is balanced by photodissociation.  The depth where the abundance reaches this maximum value depends on the strength of the UV field, the local density, and grain opacity \citep{hkbm08}.
At lower depths photodissociation will destroy any desorbed H$_2$O molecules and erode the mantle.     
At greater depths, the lack of UV photons leaves the ice mantle intact.   Interior of snow line the water ice can evaporate from the grains and also be produced in the gas via the high temperature chemistry discussed in \S~\ref{sec:gas}.   There is also a strong possibility of rapid movement of icy solids from the outer nebula that can seed additional water and other ices to inside the snow-line \citep{cc06}.

Radially there exist large gradients in the gas/solid ice ratio of molecular species in the midplane.  
 The snow-line is species specific in the sense that CO should be present in the gas phase at greater radii than water (see Fig.~\ref{fig:tsub}).   This may not be a continuous transition with various species sublimating according to their molecular properties.   Rather some molecules may be frozen in the water lattice (possibly clathrates in denser regions) and may evaporate along with water as seen in laboratory experiments \citep{sa93, collings_lab, viti_evap}.    In the dense interior a wide range of high temperature kinetic reactions are likely active.  These reactions can produce observed species, such as HCN and C$_2$H$_2$ \citep{gail02, acg08}, and perhaps beyond.
However, the eventual products are tempered by the destructive influence of photons.   As noted in \S~\ref{sec:valid},  thermodynamic equilibrium can potentially be reached in the midplane of the inner few AU and in giant planet subnebula, with the removal of some elements from the gas as solids condense.   However, the surface layers in the inner disk are dominated by the short timescale effects of stellar radiation.

All of these processes must be viewed in light of the likelihood of dynamical mixing both radially and vertically.   Models including mixing generally show that the vertical structure illustrated in Fig.~\ref{fig:overview}  is preserved with some widening of the warm molecular layer \citep{willacy06}.  Beyond the inward advection of material the outward diffusion of hot gas from the inner nebula can bring gas to cold layers where molecules  freeze onto grains surfaces \citep{tg07}. 
Moreover, the grains are evolving via coagulation and settling and the gas gradually becomes transparent to destructive radiation.   Grain evolution ultimately results in the formation of planetesimals and planets (likely at different timescales for giant and terrestrial worlds)  and the gas chemistry will continue until the gas disk dissipates.

\section{Summary}

We are approaching an age where studies of extra-solar systems can be informed by, and inform, studies of the chemical composition of bodies in our solar system.  Some major gaps remain in our understanding that will benefit from this closer cooperation.
These include (1) What is the relative importance of thermodynamical equilibrium and kinetics in the inner disk; how does this inform the meteoritic inventory.   
(2) Inclusive of pre-solar grains, does any material, remain pristine and chemically unaltered from its origin
in the parent molecular cloud?  It is likely that the deuterium enrichments ultimately originate in the cold pre-stellar stages, but are these components dissociated and re-assembled into different form?
(3) How deep inside the disk (radially) do cosmic rays penetrate and how might this influence dynamical evolution?   (4) How extensive is the complex chemistry in the inner disk and how are these organics incorporated into planetesimals?   (5) Diffusive mixing can be important, but how much and for how long is it active?    These are just a sample of the many questions that remain.
 In this chapter we have outlined some of the basic physical and chemical processes that have laid a foundation for much of our current understanding.  
   We hope that this chapter, and book, informs the next generation of researchers  in order to untangle long-standing questions regarding 
the initial conditions, chemistry, and dynamics of planet formation, 
the origin of cometary ices, and, ultimately, a greater understanding of
the organic content of gas/solid reservoirs that produced life  at least
once in the Galaxy. 

\noindent {\bf Acknowledgements}

This material is based upon work supported by the National Science Foundation under Grant No. 0707777.
The author is also grateful to the referee, Ewine van Dishoeck, and Joanna Brown for numerous helpful comments.


\bibliography{bergin}

\printindex
\end{document}